\documentclass[11pt,letterpaper,]{article}
\usepackage[T1]{fontenc}
\usepackage[utf8]{inputenc}
\usepackage{lmodern}
\usepackage{upgreek}
\usepackage{textcomp}
\usepackage{amsmath,amsfonts,amssymb,amsthm,mathrsfs}
\usepackage{latexsym}
\usepackage{authblk}
\usepackage{graphicx}
\usepackage[english]{babel}
\usepackage{xcolor}
\usepackage{soul}
\usepackage[normalem]{ulem}
\usepackage[outercaption]{sidecap}
\definecolor{apgreen}{rgb}{0.55, 0.71, 0.0}
\usepackage[top=2cm, bottom=2cm, left=2cm, right=2cm]{geometry}

\newcommand\dsout{\bgroup\markoverwith{\textcolor{orange}{\rule[0.5ex]{1pt}{1pt}}}\ULon} 
\newcommand\msout{\bgroup\markoverwith{\textcolor{red}{\rule[0.5ex]{1pt}{1pt}}}\ULon} 
\newcommand\tpout{\bgroup\markoverwith{\textcolor{blue}{\rule[0.5ex]{1pt}{1pt}}}\ULon} 
\newcommand\spout{\bgroup\markoverwith{\textcolor{apgreen}{\rule[0.5ex]{1pt}{1pt}}}\ULon} 
\newcommand\flast{\bgroup\markoverwith{\textcolor{violet}{\rule[0.5ex]{1pt}{1pt}}}\ULon} 
\newcommand\SAout{\bgroup\markoverwith{\textcolor{purple}{\rule[0.5ex]{1pt}{1pt}}}\ULon} 
\newcommand\rodout{\bgroup\markoverwith{\textcolor{cyan}{\rule[0.5ex]{1pt}{1pt}}}\ULon} 

\usepackage[backgroundcolor=yellow!30,textsize=scriptsize]{todonotes}

\usepackage{abstract}
\addto{\captionsenglish}{}

\makeatletter
\renewcommand\@biblabel[1]{#1.} 
\makeatother

\usepackage{xr}
\usepackage{setspace}
\usepackage{filecontents}

\title{\bf \sffamily Diversity patterns and speciation processes in \\a two-island system with continuous migration}
\author[1]{Débora Princepe}
\author[2]{Simone Czarnobai}
\author[1]{Thiago M. Pradella}
\author[3]{Rodrigo A. Caetano}
\author[1,4]{Flavia M. D. Marquitti}
\author[1]{Marcus A. M. de Aguiar}
\author[3]{Sabrina B. L. Araujo}
\affil[1]{Instituto de F\'{i}sica `Gleb Wataghin', Universidade Estadual de Campinas, Campinas, Brazil}
\affil[2]{Programa de Pós Graduação em Ecologia e Conservação, Universidade Federal do Paraná, Curitiba, Brazil}
\affil[3]{Departamento de Física, Universidade Federal do Paraná, Curitiba, Brazil}
\affil[4]{ Instituto de Biologia, Universidade Estadual de Campinas, Campinas, Brazil}
\date{}

\emergencystretch=\maxdimen
\begin{document}
\doublespacing 
\selectlanguage{english}
\maketitle

\begin{abstract}

Geographic isolation is a central mechanism of speciation, but perfect isolation of populations is rare. Although speciation can be hindered if gene flow is large, intermediate levels of migration can enhance speciation by introducing genetic novelty in the semi-isolated populations or founding small communities of migrants. Here we consider a two-island neutral model of speciation with continuous migration and study diversity patterns as a function of the migration probability, population size, and number of genes involved in reproductive isolation  (dubbed as genome size). For small genomes, low levels of migration induce speciation on the islands that otherwise would not occur. Diversity, however, drops sharply to a single species inhabiting both islands as the migration probability increases. For large genomes, sympatric speciation occurs even when the islands are strictly isolated. Then species richness per island increases with the probability of migration, but the total number of species decreases as they become cosmopolitan. For each genome size, there is an optimal migration intensity for each population size that maximizes the number of species. We discuss the observed modes of speciation induced by migration and how they increase species richness in the insular system while promoting asymmetry between the islands and hindering endemism.


\end{abstract}

\setlength{\parindent}{0.5cm}
\pagebreak

\section{Introduction}

Geography plays a central role in speciation. The isolation of populations imposed by geographic barriers,  allopatry,  is indeed the most straightforward process of diversification  \cite{coyne_speciation_2004,fitzpatrick2009pattern}: when isolation is complete, gene flow is interrupted, and mutations accumulated in individuals of one group are not shared with individuals of the other group, increasing the genetic discrepancies between the populations and eventually leading to reproductive isolation.  Speciation in the presence of gene flow, however, is also very frequent, occurring when isolation is partial or even when geographic barriers are completely absent \cite{gavrilets2003perspective,Hey2006,Smadja2011}.  For instance,  speciation with restricted gene flow in spatially structured populations, called parapatric, has been studied in different contexts  \cite{gavrilets_rapid_1998,gavrilets2000waiting,Gavrilets2000_patterns,aguiar_global_2009,yamaguchi2017parapatric}. Likewise, populations inhabiting a single geographic area, with no restriction to gene flow, can also split into different species, a process termed sympatric speciation.  The possibility of speciation in sympatry has been theoretically demonstrated for populations under disruptive selection \cite{smith1966sympatric,Gavrilets_2006,bolnick2007sympatric}, strong competition \cite{Dieckmann_1999}, mating preference \cite{Caetano2020} and even in neutral scenarios \cite{higgs_stochastic_1991}, but its occurrence in nature is rare and still controversial \cite{fitzpatrick2009pattern,Bolnick_2004}.

Migration between groups is a common behavior that prevents complete isolation of populations \cite{winker2000migration, nosil2008, chaine2012dispersal, Turbek2018}. For instance, estuarine-river environments and tide pools are systems under pulse-driven biotic events \cite{halas2005historical}. These environments are cyclically isolated and expanded as a consequence of the water levels \cite{baggio2017opportunity}. Thus, the system behaves like islands subject to periodic exchanges of migrants from neighboring sites. Similarly, species that have seasonal migration behavior are constantly mixing in a common area nesting or breeding place and irradiating after to other sites \cite{Cooper2017, Manthey2020, Winker2006, Everson2019}. In this case, the migration itself (associated with dispersal abilities and spatial structure of the landscape) plays the role of an intermittent geographic barrier \cite{Hey2006, Claramunt2012, Agnarsson2014, pinheiro2014rock, Linck2019, Ashby2020}. Such intermediate geographic situations can affect the process of speciation in complex ways since diversification in the presence of migration depends on a balance between colonization, local selection, and the homogenizing effects of gene flow \cite{garant2007multifarious}. On the one hand, high levels of migration can constrain diversification via maintenance of gene flow \cite{Mayr1963-MAYASA,Ronn2016} and hinder speciation \cite{fitzpatrick2009pattern, Claramunt2012}. It has been suggested that even one or a few migrants should be sufficient to avoid genetic diversification altogether \cite{slatkin1987gene}. On the other hand, migration between partially isolated populations can foster speciation by increasing genetic variation \cite{Smadja2011,Cowie2006} and by promoting founder events \cite{Spurgin2014}, which is the establishment of a small number of migrants in a new area, favoring rapid genetic changes and eventual diversification into a new species \cite{Barton1984,templeton1980theory,gavrilets1996founder, Templeton2008}. Understanding how migration increases diversity in semi-isolated populations has become increasingly important in a world of fragmented environments \cite{hagen2012biodiversity, Mills1996}.

Investigation of the role of migration on speciation has mainly focused on the time to speciation and the influence of population structure on the rate of species creation, with evolution essentially driven by mutation and genetic drift \cite{Manzo_Peliti_1994,Yamaguchi-2013_first}. Notably, migration increases the time for speciation \cite{gavrilets2000waiting}, but species divergence can be favored by subdivision of the population, even in the absence of local adaptation \cite{gavrilets_rapid_1998,Gavrilets2000_patterns}. When considering isolated populations under rare but recurrent migration, optimal intermediate migration intensities maximize the rate of species creation in two-islands systems \cite{Yamaguchi-2013_first, yamaguchi2017tipping, Yamaguchi2016_smallness}. Still, some patterns are yet to be addressed and discussed in more detail, such as the distribution of species richness, endemism, and species persistence. In this aspect, models can help elucidate and differentiate processes from patterns. Furthermore, current advances in obtaining genome-wide datasets provide robust estimates of gene flow and demographic history \cite{Hey2010, Gyllenhaal2020}, evincing that speciation with gene flow in insular systems is common and may rise even in the absence of ecological divergence \cite{Gyllenhaal2020}. Therefore, understanding how migration and diversification are related can provide a better interpretation of speciation events in these intermediary geographic configurations that are not fully sympatric or allopatric.

In the present study, we focus on how migration intensity affects speciation and species richness in insular communities. We consider a neutral model for two islands subject to frequent exchange of individuals, where mating occurs locally and is only restricted by genetic similarity. 
The model is based on the work of Manzo \& Peliti \cite{Manzo_Peliti_1994}, who considered a similar dynamic process of local reproduction and migration between two islands in the limit of infinitely large genomes, and investigated the possibility of allopatric speciation in the presence of gene flow. Here we show that population size and genetic features, particularly the number of loci involved in reproductive isolation, are key to determine not only the total number of species in the system but also the fraction of species that are endemic or cosmopolitan. Our simulations show that migration promotes the appearance of new species through two distinct processes: founding populations and induced sympatric speciation. In the first case, the small but continuous flow of migrants creates a sub-population that differentiates from the residents. In the second case, the migrants mix with the residents, creating a genetically diverse population that eventually splits into species.

\section{Methods}

Our model for insular populations is based on the theory proposed by Manzo \& Peliti \cite{Manzo_Peliti_1994}, which is itself based on the sympatric speciation model by Higgs \& Derrida \cite{higgs_stochastic_1991}. In these models, each individual is characterized by a string of infinite biallelic genes that represents the loci involved in reproductive isolation, dubbed as the individual's genome. Using  infinite genes allows the derivation of several analytic results by mean-field calculations. However, important features of the infinite genes model disappear if the genome size is small, including the possibility of sympatric speciation \cite{aguiar_speciation_2017}. Here we review the model of speciation in islands as proposed by Manzo \& Peliti and extend it to finite genomes \cite{aguiar_global_2009,Costa2018}. We shall see that genome length plays a key role in the diversification process and that the limit of infinite genes is a very particular case.

\subsection{Model overview}
\label{overview}

We consider two identical spatial regions (hereafter called islands) occupied by populations that can migrate and evolve through mutations and recombination. Each individual $\alpha$ is represented by a binary string of independent $B$ genes, $\{S_1^\alpha, S_2^\alpha, \dots, S_B^\alpha\}$, where each locus $S_i^\alpha$ can assume the alleles $\pm 1$. Individuals are hermaphroditic and reproduction is sexual. Population size is kept constant at the carrying capacity $M$ in each island, except during the migration period when small fluctuations are allowed. The dynamics starts with $2M$ clonal individuals ($M$ on each island) and goes through cyclical events of migration, reproduction, and species identification (see Fig. \ref{fig:model_diag}), as detailed below:

\begin{figure}[ht]
\begin{center}
    \includegraphics[width=0.7\linewidth]{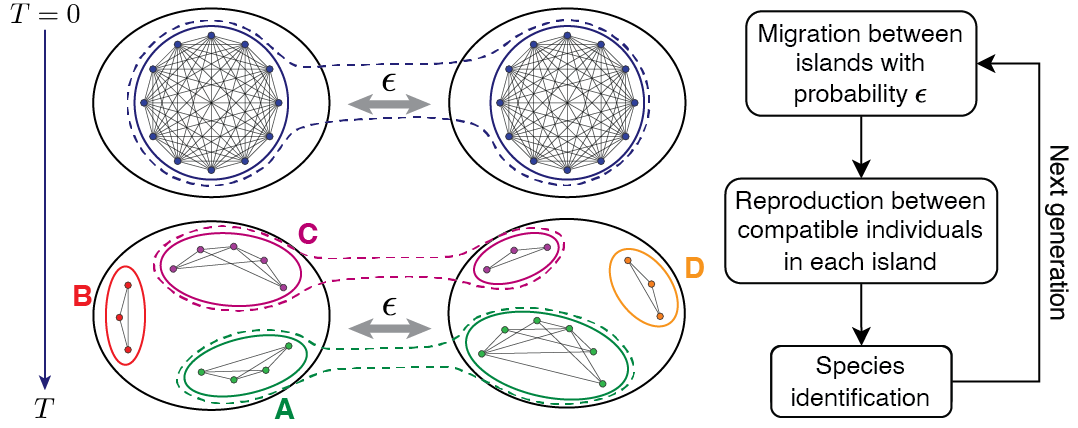}
\end{center}
 \caption{Schematic of the model dynamics. We consider two insular populations of initially identical individuals, represented by dots. At each time step, the islands exchange individuals with a probability $\epsilon$. Generations are non-overlapping and sexual reproduction occurs between individuals in the same island having a minimal genetic similarity. Species are identified at the end of each cycle based on the possibility of gene flow, illustrated by links and enclosed in ellipses. The ellipses A, B, C and D (full lines) correspond to species classification within each island. Species can be exclusive to an island (B and D), or can be common to both (A and C, dashed lines).}
 \label{fig:model_diag}
\end{figure}

{\it Migration:} at each generation, each individual has a probability $\epsilon$ of migrating to the other island. Thus the population size in each island can vary right after migration, but the total remains $2M$ individuals. We refer to this process as continuous migration, as opposed to rare migrations of larger groups \cite{Yamaguchi-2013_first}.

{\it Reproduction:} reproduction occurs after migration only between individuals on the same island. On each island, $M$ offspring are born and replace the previous population, compensating for fluctuations in populations size that may have occurred after migration. Generations do not overlap, and offspring are generated as follows: a first parent $P_1$ (focal parent) is chosen at random. A second parent $P_2$ is selected, also at random, among the remaining individuals. Reproduction between these two individuals will only occur if they are genetically compatible, i.e., if they have a minimal genetic similarity, as defined below. If the selected $P_2$ is not compatible with $P_1$, another second parent is randomly selected until this condition is met. If after $M$ trials no such individual is found, $P_1$ is discarded and a new first parent is selected. The offspring inherits, gene by gene, the allele of either parent with equal probability. The offspring's genome is also subjected to a mutation rate $\mu$ per loci, which can be reversed during the dynamics.

Reproduction is restricted by the genetic compatibility of the mating pair. Two individuals $\alpha$ and $\beta$ can reproduce only if their genetic similarity, defined by
\begin{equation}
q^{\alpha \beta} = \frac{1}{B} \sum_{i=1}^B S_i^\alpha S_i^\beta,
\label{qdef}
\end{equation}
satisfies the condition $q^{\alpha \beta} \geq q_{min}$, where the minimum similarity $q_{min}$ is a model parameter. If the genomes are identical then $q^{\alpha \beta}=1$,  while two genomes with random entries have $q^{\alpha \beta}$ close to zero. The model is neutral in the sense that there is no fitness assigned to the individuals, and the choice of mating partner is random, limited only by a minimum genetic similarity, i.e., a weak form of assortative mating.

{\it Species identification:} speciation occurs when the gene flow between groups of individuals is disrupted. The population maps to a network where the individuals (nodes) are connected if they are genetically compatible. Species correspond to the components of the network. Not all individuals of a species need to be compatible since the genetic flow can be established through intermediary individuals. In network terms, there is a path connecting them within the component, as illustrated in Figure \ref{fig:model_diag}. We identify species considering local and global classifications. In the first case, only individuals from the same island are considered (components circled by continuous lines in Fig. \ref{fig:model_diag}), while in the second, the genetic compatibility network is built regardless of which island individuals belong (components circled by dashed lines in Fig. \ref{fig:model_diag}). 

\subsection{Analytical results}

In the limit where genome size becomes infinite, the entire dynamics can be obtained by simply updating the similarity matrix. A number of simplifications occur and analytical results can be obtained \cite{higgs_stochastic_1991}, as we summarize below. 

\subsubsection{One island model}

If there is no restriction on mating, $q_{min}=0$, the overlaps $q^{\alpha \beta}$ are all very similar. The distribution of overlaps is a peaked function whose average value satisfies the evolution equation \cite{higgs_stochastic_1991} (see section \ref{app1} of the Supporting Material for a derivation)
\begin{equation}
	q_{t+1} = e^{-4\mu} \left[ \left(1-\frac{1}{M}\right)q_t + \frac{1}{M}  \right] \equiv e^{-4\mu} Q(q_t,M)
	\label{dyn1p}
\end{equation}
where $t$ is measured in generations. Setting $q_{t+1}=q_t=\bar{q}$, we find the equilibrium $\bar{q}  \equiv q_0 \approx (1+4\mu M)^{-1}$. If $q_{min} > q_0$, the distribution of similarities cannot reach the equilibrium and the population splits into species, formed by groups of individuals whose average similarity is larger than $q_{min}$. Inter-species similarities are smaller than $q_{min}$, tending to zero with time \cite{higgs_stochastic_1991}. Average species abundance $m$ can be estimated by setting $q_0 \rightarrow q_{min}$ and $M\rightarrow m$, as this gives the population size that would naturally equilibrate at $q_{min}$. We obtain $m = ({q^{-1}_{min}} - 1)/(4 \mu) $ and the average number of species is  $N_{(one)} = M/m = 4 M \mu/(q^{-1}_{min} - 1)$. We find numerically that the distribution of abundances has an average slightly smaller than $m$, which implies average number of species slightly larger than $N_{(one)}$.

We note that this dynamic behavior can change drastically for finite genomes. If the number of loci $B$ is small enough, the distribution of similarities becomes wider and, instead of splitting into separate peaks as it crosses $q_{min}$, it equilibrates as a single peak around $q_{min}$, preventing speciation. Only for $B$ above a critical value, which depends on population size, mutation rate and $q_{min}$, the above description (and therefore sympatric speciation) is recovered \cite{aguiar_speciation_2017}.

\subsubsection{Two-islands model}

In the case of two islands, we need to distinguish the similarities between individuals inhabiting the same island, $q$, or different islands, $p$. If individuals  $\alpha$ and $\beta$ were born on the same island, then their average $q$ similarity in the next generation, before migration, is given by Eq. \ref{dyn1p}. If, on the other hand, they were born in different islands, then their $p$ similarity before migration is $p_{t+1}=e^{-4\mu} p_t$ as they cannot have a common parent \cite{Manzo_Peliti_1994} (see Eq. 3 of Supplementary Material). The dynamics of $q$ and $p$ after migration is, therefore, given by the following equations:
\begin{eqnarray}
	q_{t+1} &=& a(\epsilon) e^{-4\mu} Q(q_t,M) + b(\epsilon) e^{-4\mu} p_t \nonumber \\
	p_{t+1} &=& b(\epsilon) e^{-4\mu} Q(q_t,M) + a(\epsilon) e^{-4\mu} p_t
\label{mp1}	
\end{eqnarray}
where $a(\epsilon) = (1-\epsilon)^2 + \epsilon^2$ is the probability that both $\alpha$ and $\beta$ did not migrate or that they both migrated, therefore keeping their original geographic relation, and $b(\epsilon) = 2 \epsilon (1-\epsilon)$ is the probability that one of them two exchanged places, altering the geographic relative position of the pair. 
The equilibrium solutions are obtained by setting $q_{t+1}=q_t$ and $p_{t+t}=p_t$. For $\epsilon$,  $\mu$ and $1/M$ all much smaller than 1, we obtain \cite{Manzo_Peliti_1994}
\begin{eqnarray}
	q_0(\nu,\sigma) &=& \frac{\nu + 2 \sigma}{2\sigma(2\nu+1) + \nu(\nu+1)} \nonumber \\
	p_0(\nu,\sigma) &=&  \frac{2 \sigma}{2\sigma(2\nu+1) + \nu(\nu+1)} 
	\label{q0p0}
\end{eqnarray}
where $\nu=4\mu M$ and $\sigma=M\epsilon$ is the average number of exchanged migrants at each generation. Notice that for $\sigma \gg \nu$, we have $p_0(\nu,\sigma)= q_0(2\nu,0)$, indicating that for a large migration intensity the two islands behave as a single island with twice the population.

Two new important results are derived here: first, the expected number of species in each island, $N$, can be estimated by replacing $q_0(\nu,\mu)$ by $q_{min}$ in Eq. \ref{q0p0}, as similarly done before, and solving for $M \rightarrow m$. The number of species is $N=M/m$:
\begin{equation}
    N =  \frac{\nu(4\sigma+\nu)}{(2\sigma+\nu)(q_{min}^{-1} -1)}.
    \label{eq:ns}
\end{equation}
Second, the total number of species in the islands, $N_T$, can also be estimated as 
\begin{equation}
    N_{T} =  N \left( 2 - \frac{p_0(\nu,\sigma)}{q_0(2\nu,0)} \right).
    \label{eq:nt}
\end{equation}
When $\sigma=0$ the islands are independent and $N_{T} = 2 N$. When migration dominates over mutations, $\sigma \gg \mu$, $N_{T} = N$.

\subsection {Numerical simulations and data analysis}

We performed simulations with both finite and infinite genomes to evaluate how migration probability ($\epsilon$), population size per island ($M$) and amount of loci involved in reproductive isolation ($B$) affects patterns of diversity in a two island system. The mutation rate per locus and the genetic threshold for reproduction were fixed for all simulations, $\mu=0.001$ and $q_{min}=0.9$, respectively. The parameters of interest were varied with the following values:  $B=1000, 2000, 3000, 10000,$ and the infinite model;  $M$ from 25 to 500 individuals; and $\epsilon$ from 0 to 0.5. 
Populations were evolved during 2000 generations, with observation of species richness at $T= 500$, $1000$, and $2000$ generations. We ran 50 simulations for each set of parameters, unless stated otherwise (see Supplementary Material, Fig. \ref{fig:Samples}).

When probing the species richness, we considered both local and global classifications (see \textit{Species identification} in \ref{overview}). In the local classification, the species richness per island were designated by $N_{L1}$ and $N_{L2}$. Under the global classification, the species richness in each island was $N_{1}$ and $N_{2}$, and the total number of species was $N_{T}$. Local and global classifications differed when individuals in one island, say island 1, could establish gene flow with two species of island 2 that would otherwise be reproductively isolated, a configuration that resembles the ring species assemblage \cite{martins_evolution_2013}. In this case these two species that contributed to $N_{L2}$ would count as one for $N_{2}$. We quantified these events, named here ``ring-like species'', by calculating $N_{ring}=(N_{L1}-N_{1})+(N_{L2}-N_{2})$ (the local classification was exclusively employed for this analysis). 

We also evaluated the asymmetry $\Delta N$ between the islands as the proportion of unbalance in species richness (under the global classification) in respect to the average, 
\begin{equation}
    \Delta N= \frac{\sqrt{(N_{1}-N_{2})^2}}{\bar{N}}, 
    \label{eq:asy}
\end{equation} 
where $\bar{N}=(N_{1}+N_{2})/2$. The diversity between the islands was measured based on the Jaccard distance index, whereas we calculate the number of exclusive species given by $(N_{T}-N_{1})+(N_{T}-N_{2})$. We call here the beta-diversity index $\beta_I$ the normalization of this quantity, defined as 
\begin{equation}
    \beta_I=\frac{2N_{T}-N_{1}-N_{2}}{N_T}.
\end{equation}
If species are exclusive to each island, $N_T=N_{1}+N_{2}$ and $\beta_I =1$, indicating maximum endemism. In the other hand, if species are all common to both islands, we have $N_T=N_{1}=N_{2}$ and $\beta_I=0$, i.e., all species are cosmopolitan.

\section{Results}

\begin{figure}[!ht]
\begin{center}
    \includegraphics[width=0.9\linewidth]{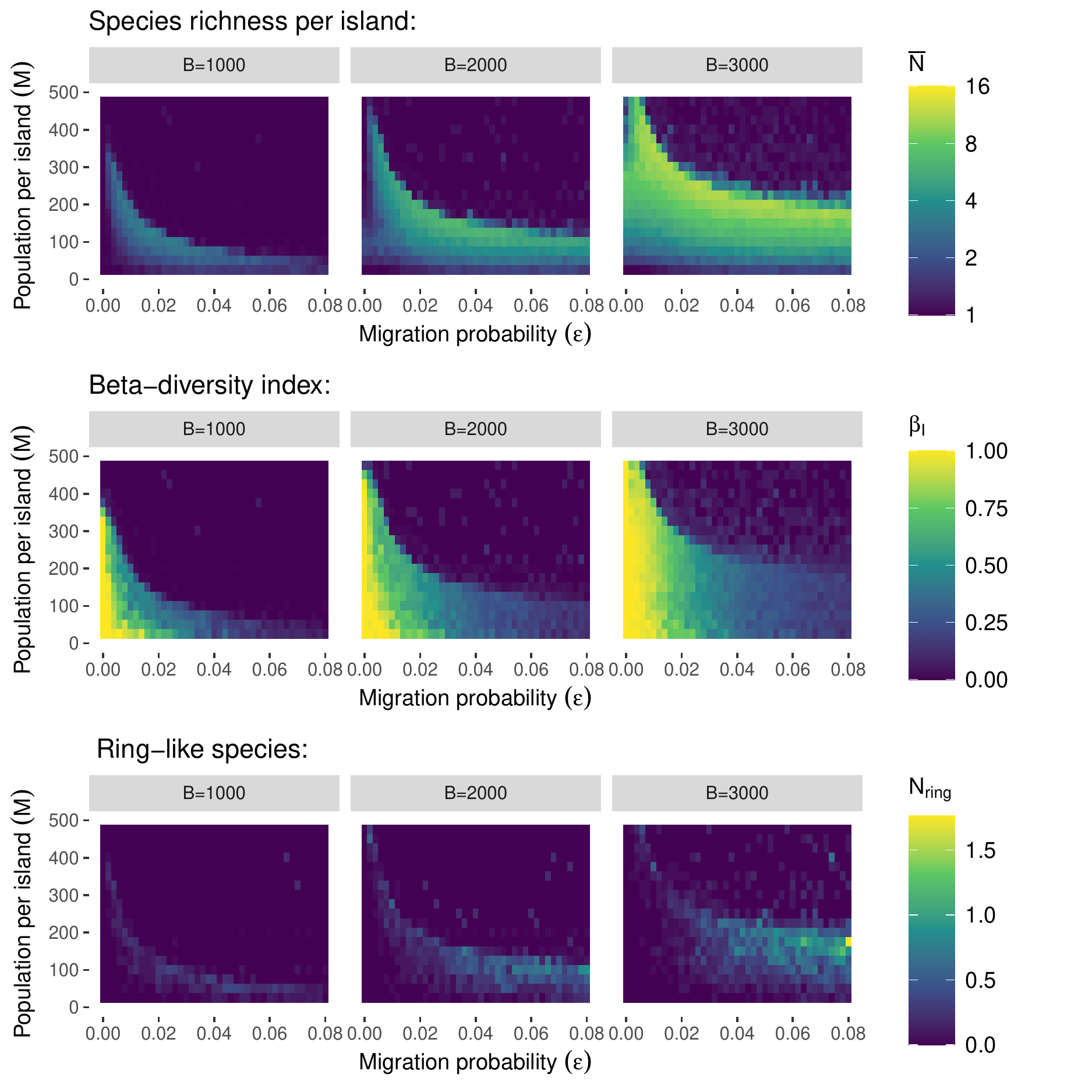}
\end{center}
  \caption{Diversity patterns in the islands as function of population size ($M$) and migration probability ($\epsilon$) for genome lengths $B=1000$, $2000$ and $3000$ at $T=2000$: average species richness per island, $\bar{N}=(N_{1}+N_{2})/2$ (first row); the proportion of species that is exclusive of an island, measured by the beta-diversity index $\beta_I$ (second row); and the number of ring-like species, $N_{ring}$ (third row). Migration increases the species richness in each island but reduces the beta-diversity, followed by the formation of ring-like species. There are optimal combinations of population size and migration probability that favor speciation for each genome length $B$. The patterns are similar through the values of $B$, with larger $B$ sustaining a higher number of species.}
  \label{fig:Diagrama_fases}
\end{figure}

We first explored how migration probability, genome size, and population size affected diversity at the end of the simulations ($T=2000$) for genome lengths $B=1000$, $2000$, and $3000$ with varying migration probability, $0\leq \epsilon \leq 0.08$. When the islands were completely isolated from each other, $\epsilon =0$, speciation did not occur within the islands for the smallest genome size, $B=1000$ (Fig. \ref{fig:Diagrama_fases}, upper left). In this case, the populations differentiated only through the accumulation of mutations in each island (allopatry), and each island had a unique exclusive species (indicated by $\beta_I=1$ in Fig. \ref{fig:Diagrama_fases}, bottom left). This inter-island divergence was limited by the population size ($M \lesssim 350$ individuals), as larger $M$ required longer times for speciation (see Fig. \ref{fig:B1000}). For $B=2000$, sympatric speciation in isolated islands was rare, occurring scarcely only for small values of $M$ (Fig. \ref{fig:Diagrama_fases}, upper center). On the other hand, for $B=3000$, isolated islands presented sympatric speciation for all $M \gtrsim 50$ (Fig. \ref{fig:Diagrama_fases}, upper right). Increasing $B$ also led to an increase in the maximum population size that supported $\beta_I=1$, the maximum endemism (Fig. \ref{fig:Diagrama_fases}, middle). Strict allopatry ($\epsilon=0$) represented the peak of endemism for all scenarios (see Fig. \ref{fig:Diagrama_fases}, middle).
 
Migration even at low levels was sufficient to increase the number of species on each island. However, the beta-diversity decreased with migration, as expected, with a slower decay the larger the genome size. Therefore, although species in each island were more numerous, they were more likely to be shared. Regardless of the genome size, a curve depending on population size and migration probability maximized the per island and total species richness, with peak values increasing with $B$ (Fig. \ref{fig:Diagrama_fases}, top; for the values of the total species richness, $N_T$, see Fig. \ref{fig:B1000}, \ref{fig:B2000} and \ref{fig:B3000}).  The increase of the average number of species per island, $\bar{N}$, was followed by the formation of ring-like species (Fig. \ref{fig:Diagrama_fases}, middle), that is, groups of individuals with disrupted gene flow in the same island but with the possibility of gene flow through individuals from the other island. The increase in $N_{ring}$ was simultaneous to the loss of endemism, indicating the genetic homogenization due to more intense migration. Those patterns were consistent over time, with a direct dependence on the time for allopatry with $M$ (see Fig. \ref{fig:B1000}, \ref{fig:B2000} and \ref{fig:B3000}).

\begin{figure}[ht]
    \includegraphics[width=\linewidth]{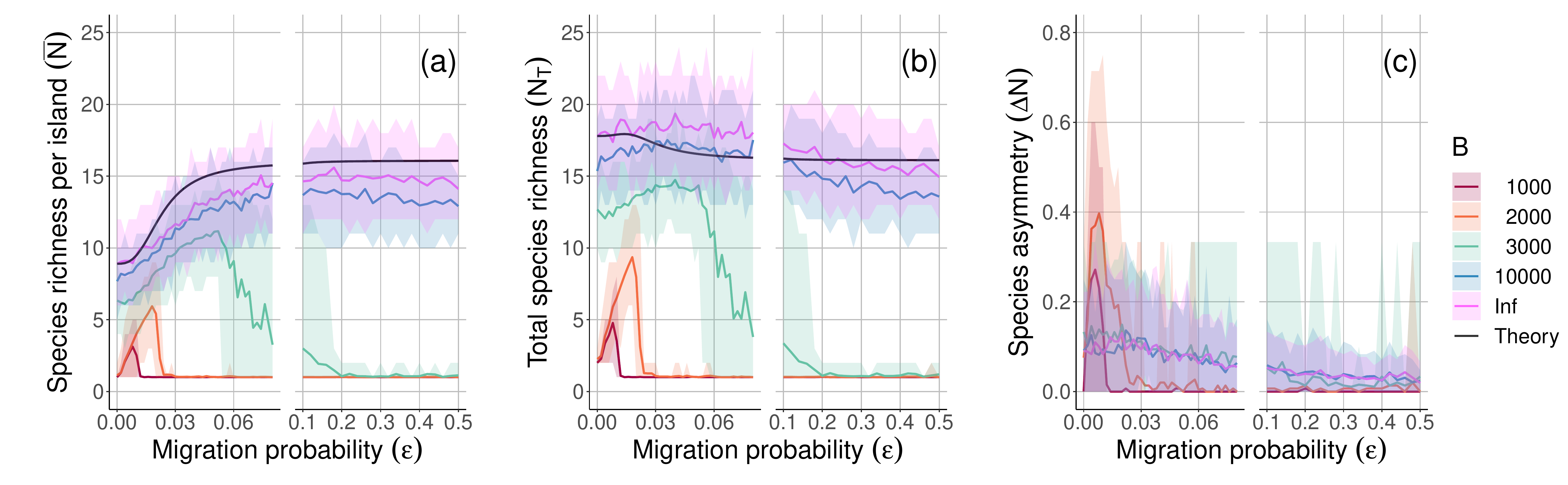}
  \caption{Species richness and asymmetry between the islands at $T=2000$ for varying migration probability ($\epsilon$) and genome sizes $B$ (colors) with fixed population size ($M=200$). Results are shown for 50 independent simulations. The solid lines represent the average value of all realizations, and the shadowed areas show a confidence interval of  90\%. The full black lines represent the expected values from Eqs. \ref{eq:ns} and \ref{eq:nt}. (a) Average number of species per island, $\bar{N}=(N_{1}+N_{2})/2$. (b) Total number of species, $N_T$. (c) Species asymmetry $\Delta N$, calculated with Eq. \ref{eq:asy}. The higher asymmetry occurs for values of $B$ for which migration is essential for speciation. }
  \label{fig:asy}
\end{figure}

We also simulated larger genomes ($B=10000$ and infinite loci) to compare with the predictions from the analytical solutions. The results were analyzed in more detail by fixing the population size at $M=200$ and varying the migration probability in the range $0\leq \epsilon \leq 0.5$. The theoretical prediction for the species richness in each island $N$ (Eq. \ref{eq:ns}), based on a model of infinite loci, was slightly lower than the average value from the simulations. It occurs because the prediction considers a uniform distribution of species abundances (constant $m$), and the simulated populations resulted, in fact, in heterogeneous species abundances. For a better comparison, we adjusted the result from Eq. \ref{eq:ns} for isolated islands ($N=7.2$ for $\epsilon=\sigma=0$) to the simulations by adding a correction ($\delta = 1.8$), and shifted the whole curve by the same constant (black line in Fig. \ref{fig:asy}a). This theoretical expectation of the number of species in one island showed a monotonic increase with $\epsilon$ and was also a reasonable estimate of the number of species for $B=10000$. For large genomes ($B\geq10000$), the population split into several species even in strict isolation (sympatric speciation), and migration increased species richness further. However, intense migration, ranging from $\epsilon=0.1$ to $0.5$, led to the homogenization of the populations that evolved effectively as contiguous $2M$ individuals. For small genomes, $B=1000$ and $2000$, migration was essential to increment species richness, as seen before. However, here we observe that this effect was limited to low levels of migration: above a critical point, the number of species within the island collapsed to a single one shared by the two islands ($\epsilon \geq  0.014$ for $B=1000$ and $\epsilon \geq 0.03$ for $B=2000$). $B=3000$ had an intermediary behavior, presenting several species for $\epsilon=0$, but also collapsed for high levels of migration ($\epsilon \geq 0.2$) with a continuous transition. The increment in total species richness ($N_T$) beyond the number of total species in strict allopatry ($\epsilon=0$) showed that migration induced some form of speciation in the insular community, and the effect was more intense for small genome lengths (Fig. \ref{fig:asy}b). The approximation for $N_T$ from Eq. \ref{eq:nt} was a reasonable estimate for $B\geq10000$, predicting the decay in $N_T$ with $\epsilon$ but in a steeper way. The qualitative patterns of species richness with respect to the migration probability when increasing $B$, keeping all other parameters fixed, are reproduced in a similar way with decreasing $q_{min}$ (Fig. \ref{fig:qmin} a) and increasing $ \mu$ (Fig. \ref{fig:qmin} b) with all other parameters fixed -- see Section \ref{SM-tradeoff} of the Supplementary Material for a discussion of the trade-off between the model parameters not explored here.

Although the migration probability was symmetric between islands, the number of species in each one was asymmetric, especially for low and intermediary migration intensities,  $0\leq\epsilon\leq0.03$ (Fig. \ref{fig:asy}c). This asymmetry occurred most of the time but was ephemeral, as we observed an alternation of the island with more species and fluctuation of the species richness on both islands (videos included in the Supporting Material). Comparing to a null model of random distribution of the species between the two islands with equal probability (see the Supplementary Material, section \ref{SM-asy}), we verified that some of the asymmetry results from inherent stochasticity of the system, but the highest asymmetry, beyond the expected by chance, occurred for values of $B$ and $\epsilon$ for which migration was the essential mechanism for the speciation. We inferred that asymmetry occurred when speciation induced by migration enhanced random imbalances between the number of species on each island. For instance, when one island was composed of a single species, migrants were likely to establish a new species on the arrival island as they were genetically compatible. On the other hand, as the other island then had more than one species, its migrants had a reduced chance of being conspecific and were more likely to go extinct on the island of arrival for lack of compatible mates, keeping the first island with a single species.

\begin{figure}[ht]
\begin{center}
    \includegraphics[width=\linewidth]{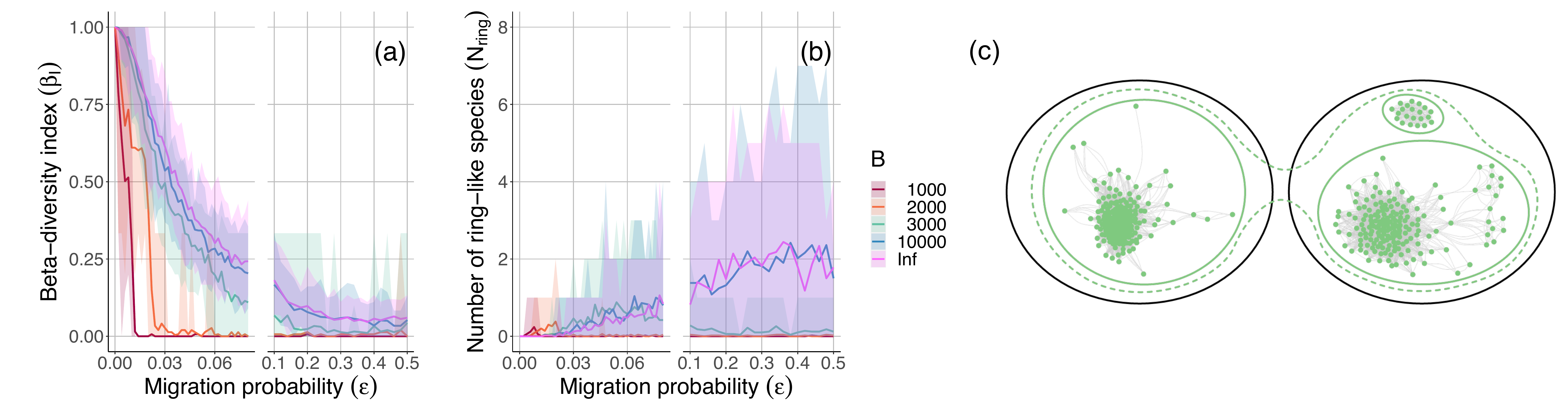}
\end{center}
 \caption{ Beta-diversity index of the insular community and ring-like species at $T=2000$ for varying migration probability ($\epsilon$) and genome sizes $B$ (colors), with fixed population size ($M=200$). (a) The beta-diversity index $\beta_I$ decreases monotonically with migration from full allopatry ($\epsilon=0$ and $\beta=1$) to the mixed populations sharing most of the species ($\epsilon \approx 0.5$ and $\beta \approx 0$). (b) As migration increases, gene flow disrupted within one island can be re-established by individuals from the other island, in a configuration similar to ring species.  (c) Example of ring-like species, for $B=2000$ and $\epsilon=0.02$ at $T=430$. Locally there is one species at the left island and two species at the right one, but, at the global classification, individuals of the first island reconnect the two clusters from the second, and then the islands share a single species.}
 \label{fig:beta}
\end{figure}

Analysis of the beta-diversity index in the insular community shows that, as expected, increasing the migration probability led to a decline in $\beta_I$ (Fig. \ref{fig:beta}a), from the complete endemism when islands were isolated ($\epsilon=0$ and $\beta_I=1$) to the mixed populations, having most of the species in common ($\epsilon \approx 0.5$ and $\beta_I \approx 0$). Large genomes were more resilient, with slow decay, while the small genomes presented a sharp drop, following the collapse in the number of per island and total species to a single one. Although the proportion of exclusive species in the system always decreased, the absolute number of exclusive species increased with migration for small genomes (Fig. \ref{fig:exclusive}). The increase in the migration probability also caused the appearance of ring-like species (Fig. \ref{fig:beta}b). Figure \ref{fig:beta}c illustrates how it occurred during the simulations. In this example, using the local referential, the island at the left had one species and the other two species, $N_{L1}=1$ and $N_{L2}=2$. 
However, the two species in the second island were, in fact, a single one in the global classification, with gene flux between the two groups reconnected by individuals of island one, then $N_{1}=N_{2}=1$. This assemblage is similar to ring species \cite{martins_evolution_2013} on a much smaller scale and longing only through a few generations, although appearing throughout the dynamics and even at $T=2000$, as shown in Fig. \ref{fig:beta}b. Such a configuration either preceded speciation (when the disrupted gene flow later led to the formation of species) or disappeared with the subsequently re-establishment of gene flow by the arrival of migrants or due to mutation. Therefore we classified them as ``ring-like species''. For small genomes, they occurred in a small number, restricted to the range of migration intensity that made $N_i>1$. For large genomes, they were recurrent and appeared in significant numbers under intense migration.

\begin{figure}[!ht]
\begin{center}
    \includegraphics[width=0.9\linewidth]{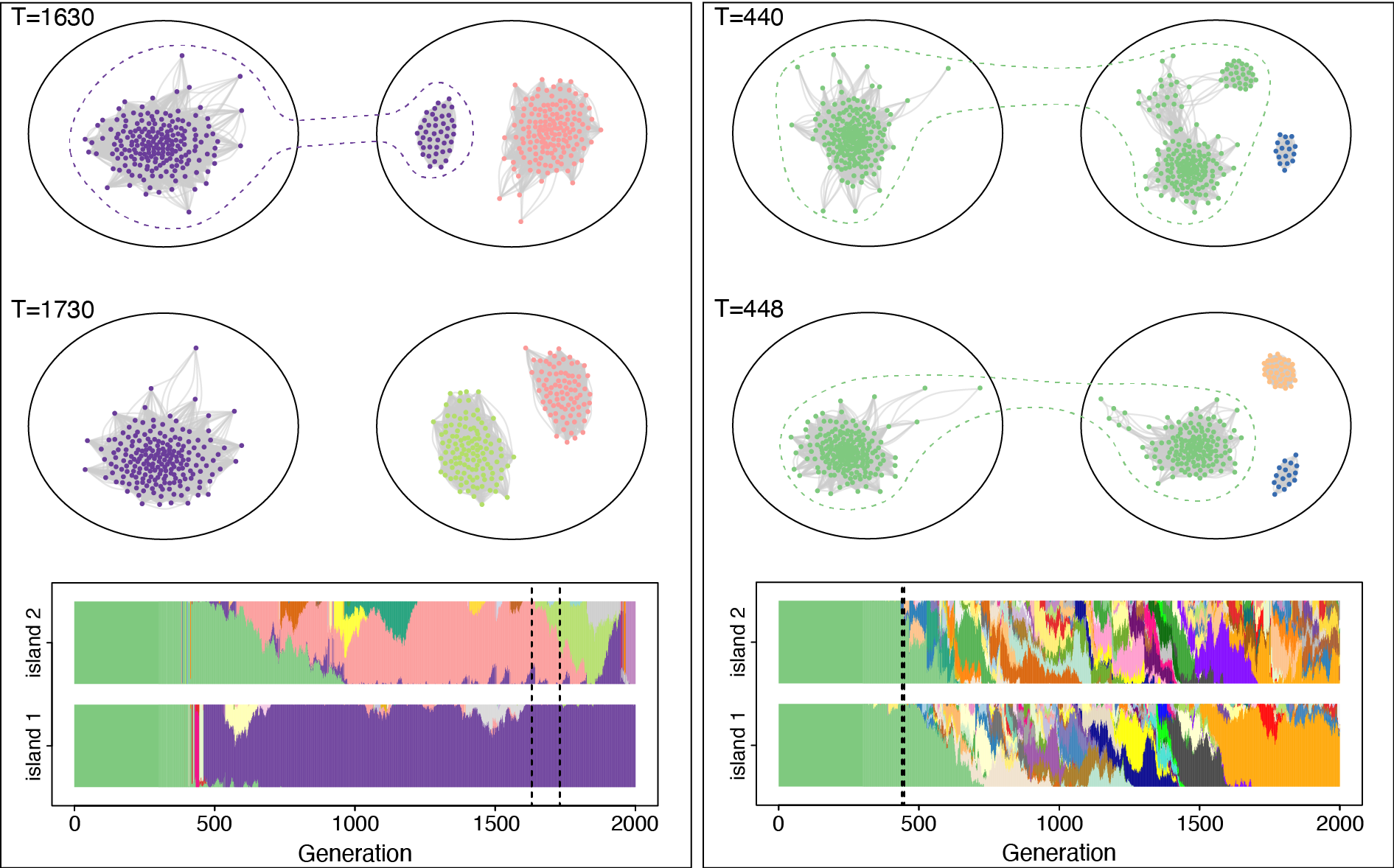}
\end{center}

 \caption{ Evolution of the populations for $B=1000$ and $\epsilon=0.004$ (left) and $B=2000$ and $\epsilon=0.02$ (right) exemplifying the speciation modes induced by migration. Bars on the bottom indicate the distribution of species abundances in each island over time. Individuals of the same species have the same color and are connected by links or dashed lines when they belong to different islands. In the speciation through founder populations (left panel), a portion of migrants differentiates into a new species without mixing with the native population. In the second mode (right panel), migrants can incorporate into the resident population, then differentiation between islands is hindered, but migration promotes sympatric speciation.
 } 
 \label{fig:dynamics}
\end{figure}

Finally, we investigated the processes by which migration leads to speciation. Using genetic compatibility networks, we observed two speciation modes induced by migration, illustrated in Figure \ref{fig:dynamics}. {\it Speciation by founding populations} occurred when migrants arriving on an island could not reproduce with the resident population for being of a different species and, when they were able to accumulate through several generations and establish a population, ended up founding a new species (Fig. \ref{fig:dynamics}, left panel). We hypothesize that this was the dominant mode for low migration probabilities based on the examples observed with videos. The second mode, recognized as a {\it sympatric speciation induced by migration} (shown in the right panel), seems to occur more frequently under slightly larger migration probabilities. Here the islands presented common species, then migrants could reproduce with the resident population, and their incorporated genetic novelties promoted speciation. The described effects can be observed in detail in the videos included with the Supporting Material.

\section{Discussion}

The present model was based on the work of Manzo \& Peliti \cite{Manzo_Peliti_1994}, who considered only genomes with infinitely many loci. They showed that when mating was not restricted by a minimum genetic similarity, i.e., $q_{min}=0$, the average values of intra and inter-island similarities ($q$ and $p$) converged to the constant values $q_0$ and $p_0$ given by Eq. \ref{q0p0}. They proposed that setting $p_0 < q_{min} < q_0$ would drive the distribution of intra-island similarities to $q_0$, avoiding sympatric speciation and at the same time creating inter-island reproductive isolation. For the parameter range studied in \cite{Manzo_Peliti_1994}, this scenario was only achieved by setting $q_{min}\approx0.2$. For finite and sufficiently small genome sizes as explored here, on the other hand, sympatric speciation can be avoided even if $q_{min} > q_0$, as the distribution of intra-island similarities becomes stationary around $q_{min}$ \cite{aguiar_speciation_2017}. The dynamics with finite genome sizes also brought up a richness of processes and outcomes that do not occur in the extreme case of infinite genomes. The time for allopatry, for example, does not depend on population size for infinite loci (Eq. \ref{eq:tallop} SM), but strongly depends on $M$ for finite genomes (Fig. \ref{fig:B1000}, \ref{fig:B2000} and \ref{fig:B3000}). Furthermore, we extended the Manzo \& Peliti model, obtaining expressions for the number of species and estimating the percentage of exclusive species on each island. Importantly, we studied the mechanisms of speciation induced by migration for infinite and finite genomes, showing how migration increases species richness while hindering endemism and promoting asymmetry between the islands and transient ring species. 

Strict allopatry with one species per island was only maintained for small genomes ($B=1000$ and $2000$) in the absence of migration with $M=200$. Even a few migrants were sufficient to promote speciation. For instance, with $\epsilon=0.004$ (equivalent to an average number of $\sigma=0.8$ migrants per generation), speciation by founding populations already occurred (see Fig. \ref{fig:dynamics} left panel and videos in the Supplementary Material). This mode was mainly related to low migration intensity and benefited from the slow dynamics of those genomes sizes (Fig. \ref{fig:Temporal}). When the two islands still had the same species, the migrants were genetically similar and, in a small number, did not bring sufficient novelties to cause a break in gene flow. After the islands differentiated from each other (around $T>400$ generations), recently arriving migrants in each island could not reproduce with the resident population and either accumulated through several generations or went extinct. When a population was established in the arrival island, they differentiated into a new species (Fig. \ref{fig:dynamics}, left panel). No sympatry was observed in those cases, in the sense that the island's resident species did not branch. Therefore, the ancestry of the new species was always fully connected to migrants, i.e., they did not share common ancestors with the former (native) species. 

With a slightly higher level of migration, the islands had more than one species and shared some of them. Although there was a loss of endemism, sympatry induced by migration was favored: migrants could reproduce with residents, and their mixing led to speciation. Remarkably, despite the higher proportion of shared species, the absolute number of exclusive species increased with migration for small genomes (Fig. \ref{fig:exclusive}). This connection between high species richness with low endemism or low species richness and high endemism in insular systems has been observed in Darwin finches, for instance \cite{Hamilton1963}, and explained by the spatial structure of the islands \cite{Gascuel2016}. Low levels of migration can be understood as a great physical distance between the islands, while a moderate level indicates islands closer to each other, establishing an analogy between our results and previous findings. Also, an intermediary level of migration seemed to optimize the evolution of diversity, which has been suggested in different contexts \cite{Yamaguchi-2013_first,garant2007multifarious}, for instance, in the intermediate dispersal model \cite{Agnarsson2014,Ashby2020}. Based on the theory of island biogeography and associating migration with dispersal abilities, it proposes that colonization promotes speciation at the same time that may increase the probability of extinction; therefore, moderate levels of migration optimize species richness in islands \cite{Agnarsson2014,Ashby2020}. Our model reproduces this effect while also providing a mechanistic view of the processes by which migration promotes speciation.

Small genome sizes were positively correlated to the asymmetry in species richness between the islands beyond the expected by chance (i.e. if the given total number of species was randomly distributed between the islands with equal probability). The effect was more significant for low levels of migration (Fig. \ref{fig:asy}c, also \ref{fig:sup_asy} and \ref{fig:Histograms}), and we hypothesized it was related to speciation by founding populations that, as mentioned before, enhanced random imbalances of species richness. When sympatric speciation took place, increasing the number of species, asymmetry was less observed. Asymmetry in the geographic range of recently branched sister species is expected under peripatric speciation and is more likely to occur in small populations \cite{Barraclough2000}. Here asymmetry in species richness resulted from gene flow in a symmetric setup (same number of individuals and migration probability in both islands) with a small number of species and not from landscape heterogeneity or different species ranges. Further investigation might address how such imbalances affect species abundances and the structure of the phylogenies.

Diversity under high migration could only be sustained with large genomes, which can be explained by the trade-off between population size and genome length (see also Fig. \ref{fig:fig2rev}). In this model, when considering a single island, sympatric speciation does not occur if the genome is too small due to the low genetic variability and the slow accumulation of mutations \cite{aguiar_speciation_2017}. The required variability is readily provided by low and moderate migration. However, under higher exchange of migrants, the populations evolve as contiguous, and then the time to speciation increases for finite genomes (see \cite{aguiar_speciation_2017} and Fig. \ref{fig:B1000}, \ref{fig:B2000} and \ref{fig:B3000}), hindering speciation. Large genomes, however, can sustain higher variability and present fast dynamics. Nevertheless, the robustness of large genomes made their response to migration less compelling.

Along the whole dynamics, we observed the formation of ring-like species, but they were more important for large genomes under high migration intensity, indicating how gene flow was actively connecting the populations. Previous models for islands with migration predicted the occurrence of ring species but in larger chains of patches \cite{gavrilets_rapid_1998}; here we find a similar effect with only two islands. However, they do not resemble the observed ring species \cite{martins_evolution_2013}, for having a much smaller extension and low durability, but can be later used as an indicative or step of the speciation process.

Finally, we note that the results presented here are conditioned to the regime of continuous migration and relatively large mutation probability ($\mu M \approx 0.1$). Rare migrations of larger groups can lead to different outcomes if the time between migrations is larger than $M$ and $\mu \ll 1/M$, allowing fixation of migrants alleles between migrations \cite{Yamaguchi2016_smallness, yamaguchi2021}. Hence, diversity in island systems depends on a large number of factors that include periodicity of migration, number of individuals, mutation probability and number of loci and alleles involved in reproductive isolation. Therefore, comparing models with data requires careful analysis of the situation at hand.

\subsection*{Code availability}
All Fortran codes used in this study are available in the GitHub repository at https://github.com/deborapr/two-islands.

\subsection*{Conflict of interest}
The authors declare that they have no conflict of interests.

\subsection*{Acknowledgment}
This work was partly supported by the São Paulo Research Foundation (FAPESP), grants \#2018/11187-8 (DP), \#2019/24449-3 (DP), \#2019/20271-5 (MAMA) and \#2016/01343-7 (ICTP-SAIFR). FMDM was supported by Coordenação de Aperfeiçoamento de Pessoal de Nível Superior -- Brazil (Finance Code 001). MAMA and SBLA were supported by Conselho Nacional de Pesquisas Científicas (CNPq), grants \#301082/2019-7 and \#311284/2021-3 respectively. SBLA acknowledge the computational support from Professor Carlos M. de Carvalho at LFTC-DFis-UFPR.

\bibliographystyle{vancouver}
\bibliography{referencias}

\begin{thebibliography}{10}

\bibitem{coyne_speciation_2004}
Coyne JA, Orr HA.
\newblock Speciation.
\newblock 1st ed. Sinauer Associates, Inc.; 2004.

\bibitem{fitzpatrick2009pattern}
Fitzpatrick B, Fordyce J, Gavrilets S.
\newblock Pattern, process and geographic modes of speciation.
\newblock Journal of Evolutionary Biology. 2009;22(11):2342--2347.

\bibitem{gavrilets2003perspective}
Gavrilets S.
\newblock Perspective: models of speciation: what have we learned in 40 years?
\newblock Evolution. 2003;57(10):2197--2215.

\bibitem{Hey2006}
Hey J.
\newblock {Recent advances in assessing gene flow between diverging populations
  and species}.
\newblock Current Opinion in Genetics {\&} Development. 2006;16(6):592--596.

\bibitem{Smadja2011}
Smadja CM, Butlin RK.
\newblock {A framework for comparing processes of speciation in the presence of
  gene flow}.
\newblock Molecular Ecology. 2011;20(24):5123--5140.

\bibitem{gavrilets_rapid_1998}
Gavrilets S, Li H, Vose MD.
\newblock Rapid parapatric speciation on holey adaptive landscapes.
\newblock Proceedings of the Royal Society B: Biological Sciences.
  1998;265(1405):1483--1489.

\bibitem{gavrilets2000waiting}
Gavrilets S.
\newblock Waiting time to parapatric speciation.
\newblock Proceedings of the Royal Society of London Series B: Biological
  Sciences. 2000;267(1461):2483--2492.

\bibitem{Gavrilets2000_patterns}
Gavrilets S, Li H, Vose MD.
\newblock {Patterns of parapatric speciation}.
\newblock Evolution. 2000;54(4):1126--1134.

\bibitem{aguiar_global_2009}
de~Aguiar MAM, Baranger M, Baptestini EM, Kaufman L, Bar-Yam Y.
\newblock {Global patterns of speciation and diversity}.
\newblock Nature. 2009;460(7253):384--387.

\bibitem{yamaguchi2017parapatric}
Yamaguchi R, Iwasa Y.
\newblock Parapatric speciation in three islands: dynamics of geographical
  configuration of allele sharing.
\newblock Royal Society Open Science. 2017;4(2):160819.

\bibitem{smith1966sympatric}
Smith JM.
\newblock Sympatric speciation.
\newblock The American Naturalist. 1966;100(916):637--650.

\bibitem{Gavrilets_2006}
Gavrilets S.
\newblock {The Maynard Smith model of sympatric speciation}.
\newblock Journal of Theoretical Biology. 2006;239(2):172--182.

\bibitem{bolnick2007sympatric}
Bolnick DI, Fitzpatrick BM.
\newblock Sympatric speciation: models and empirical evidence.
\newblock Annual Review of Ecology, Evolution, and Systematics.
  2007;38:459--487.

\bibitem{Dieckmann_1999}
Dieckmann U, Doebeli M.
\newblock On the origin of species by sympatric speciation.
\newblock Nature. 1999;400(6742):354--357.

\bibitem{Caetano2020}
Caetano RA, Sanch{\'{e}}z S, Costa CLN, {De Aguiar} MAM.
\newblock {Sympatric speciation based on pure assortative mating}.
\newblock Journal of Physics A: Mathematical and Theoretical.
  2020;53(15):155601.

\bibitem{higgs_stochastic_1991}
Higgs PG, Derrida B.
\newblock Stochastic models for species formation in evolving populations.
\newblock Journal of Physics A: Mathematical and General.
  1991;24(17):L985--L991.

\bibitem{Bolnick_2004}
Bolnick DI.
\newblock Waiting for Sympatric Speciation.
\newblock Evolution. 2004;58(4):895--899.

\bibitem{winker2000migration}
Winker K.
\newblock Migration and speciation.
\newblock Nature. 2000;404(6773):36--36.

\bibitem{nosil2008}
Nosil P.
\newblock Speciation with gene flow could be common.
\newblock Molecular Ecology. 2008;17(9):2103--2106.

\bibitem{chaine2012dispersal}
Chaine AS, Clobert J.
\newblock Dispersal.
\newblock In: Behavioural responses to a changing world. Oxford University
  Press; 2012. .

\bibitem{Turbek2018}
Turbek SP, Scordato ESC, Safran RJ.
\newblock {The Role of Seasonal Migration in Population Divergence and
  Reproductive Isolation}.
\newblock Trends in Ecology {\&} Evolution. 2018;33(3):164--175.

\bibitem{halas2005historical}
Halas D, Zamparo D, Brooks DR.
\newblock A historical biogeographical protocol for studying biotic
  diversification by taxon pulses.
\newblock Journal of Biogeography. 2005;32(2):249--260.

\bibitem{baggio2017opportunity}
Baggio RA, Stoiev SB, Spach HL, Boeger WA.
\newblock Opportunity and taxon pulse: the central influence of coastal
  geomorphology on genetic diversification and endemism of strict estuarine
  species.
\newblock Journal of Biogeography. 2017;44(7):1626--1639.

\bibitem{Cooper2017}
Cooper EA, Uy JAC.
\newblock {Genomic evidence for convergent evolution of a key trait underlying
  divergence in island birds}.
\newblock Molecular Ecology. 2017;26(14):3760--3774.

\bibitem{Manthey2020}
Manthey JD, Oliveros CH, Andersen MJ, Filardi CE, Moyle RG.
\newblock {Gene flow and rapid differentiation characterize a rapid insular
  radiation in the southwest Pacific (Aves: Zosterops)}.
\newblock Evolution. 2020;74(8):1788--1803.

\bibitem{Winker2006}
Winker K, Pruett CL.
\newblock {Seasonal Migration, Speciation, and Morphological Convergence in the
  Genus Catharus (Turdidae)}.
\newblock The Auk. 2006;123(4):1052--1068.

\bibitem{Everson2019}
Everson KM, McLaughlin JF, Cato IA, Evans MM, Gastaldi AR, Mills KK, et~al.
\newblock {Speciation, gene flow, and seasonal migration in Catharus thrushes
  (Aves:Turdidae)}.
\newblock Molecular Phylogenetics and Evolution. 2019;139:106564.

\bibitem{Claramunt2012}
Claramunt S, Derryberry EP, {J  V  Remsen} J, Brumfield RT, Remsen JV,
  Brumfield RT.
\newblock {High dispersal ability inhibits speciation in a continental
  radiation of passerine birds}.
\newblock Proceedings of the Royal Society B: Biological Sciences.
  2012;279(1733):1567--1574.

\bibitem{Agnarsson2014}
Agnarsson I, Cheng RC, Kuntner M.
\newblock {A Multi-Clade Test Supports the Intermediate Dispersal Model of
  Biogeography}.
\newblock PLOS ONE. 2014;9(1):e86780.

\bibitem{pinheiro2014rock}
Pinheiro F, Cozzolino S, Draper D, de~Barros F, F{\'e}lix LP, Fay MF, et~al.
\newblock {Rock outcrop orchids reveal the genetic connectivity and diversity
  of inselbergs of northeastern Brazil}.
\newblock BMC Evolutionary Biology. 2014;14(1):1--16.

\bibitem{Linck2019}
Linck E, Battey CJ.
\newblock {On the relative ease of speciation with periodic gene flow}.
\newblock bioRxiv. 2019;p. 758664.

\bibitem{Ashby2020}
Ashby B, Shaw AK, Kokko H.
\newblock {An inordinate fondness for species with intermediate dispersal
  abilities}.
\newblock Oikos. 2020;129(3):311--319.

\bibitem{garant2007multifarious}
Garant D, Forde SE, Hendry AP.
\newblock The multifarious effects of dispersal and gene flow on contemporary
  adaptation.
\newblock Functional Ecology. 2007;21(3):434--443.

\bibitem{Mayr1963-MAYASA}
Mayr E.
\newblock Animal Species and Evolution.
\newblock Belknap of Harvard University Press; 1963.

\bibitem{Ronn2016}
von R{\"{o}}nn JAC, Shafer ABA, Wolf JBW.
\newblock {Disruptive selection without genome-wide evolution across a
  migratory divide}.
\newblock Molecular Ecology. 2016;25(11):2529--2541.

\bibitem{slatkin1987gene}
Slatkin M.
\newblock Gene flow and the geographic structure of natural populations.
\newblock Science. 1987;236(4803):787--792.

\bibitem{Cowie2006}
Cowie RH, Holland BS.
\newblock {Dispersal is fundamental to biogeography and the evolution of
  biodiversity on oceanic islands}.
\newblock Journal of Biogeography. 2006;33(2):193--198.

\bibitem{Spurgin2014}
Spurgin LG, Illera JC, Jorgensen TH, Dawson DA, Richardson DS.
\newblock {Genetic and phenotypic divergence in an island bird: Isolation by
  distance, by colonization or by adaptation?}
\newblock Molecular Ecology. 2014;23(5):1028--1039.

\bibitem{Barton1984}
Barton NH, Charlesworth B.
\newblock {Genetic revolutions, founder effects, and speciation}.
\newblock Annual Review of Ecology and Systematics. 1984;15(1):133--164.

\bibitem{templeton1980theory}
Templeton AR.
\newblock The theory of speciation via the founder principle.
\newblock Genetics. 1980;94(4):1011--1038.

\bibitem{gavrilets1996founder}
Gavrilets S, Hastings A.
\newblock Founder effect speciation: a theoretical reassessment.
\newblock The American Naturalist. 1996;147(3):466--491.

\bibitem{Templeton2008}
Templeton AR.
\newblock {The reality and importance of founder speciation in evolution}.
\newblock BioEssays. 2008;30(5):470--479.

\bibitem{hagen2012biodiversity}
Hagen M, Kissling WD, Rasmussen C, De~Aguiar MA, Brown LE, Carstensen DW,
  et~al.
\newblock Biodiversity, species interactions and ecological networks in a
  fragmented world.
\newblock Advances in Ecological Research. 2012;46:89--210.

\bibitem{Mills1996}
Mills LS, Allendorf FW.
\newblock {The One-Migrant-per-Generation Rule in Conservation and Management}.
\newblock Conservation Biology. 1996;10(6):1509--1518.

\bibitem{Manzo_Peliti_1994}
Manzo F, Peliti L.
\newblock Geographic speciation in the {Derrida-Higgs} model of species
  formation.
\newblock Journal of Physics A: Mathematical and General. 1994;27(21):7079.

\bibitem{Yamaguchi-2013_first}
Yamaguchi R, Iwasa Y.
\newblock First passage time to allopatric speciation.
\newblock Interface Focus. 2013;3(6):20130026.

\bibitem{yamaguchi2017tipping}
Yamaguchi R, Iwasa Y.
\newblock A tipping point in parapatric speciation.
\newblock Journal of Theoretical Biology. 2017;421:81--92.

\bibitem{Yamaguchi2016_smallness}
Yamaguchi R, Iwasa Y.
\newblock {Smallness of the number of incompatibility loci can facilitate
  parapatric speciation}.
\newblock Journal of Theoretical Biology. 2016;405:36--45.

\bibitem{Hey2010}
Hey J.
\newblock {Isolation with Migration Models for More Than Two Populations}.
\newblock Molecular Biology and Evolution. 2010;27(4):905--920.

\bibitem{Gyllenhaal2020}
Gyllenhaal EF, Mapel XM, Naikatini A, Moyle RG, Andersen MJ.
\newblock {A test of island biogeographic theory applied to estimates of gene
  flow in a Fijian bird is largely consistent with neutral expectations}.
\newblock Molecular Ecology. 2020;29(21):4059--4073.

\bibitem{aguiar_speciation_2017}
de~Aguiar MAM.
\newblock {Speciation in the Derrida-Higgs model with finite genomes and
  spatial populations}.
\newblock Journal of Physics A: Mathematical and Theoretical.
  2017;50(8):085602.

\bibitem{Costa2018}
Costa CLN, Lemos-Costa P, Marquitti FMD, Fernandes LD, Ramos MF, Schneider DM,
  et~al.
\newblock {Signatures of Microevolutionary Processes in Phylogenetic Patterns}.
\newblock Systematic Biology. 2018;68(1):131--144.

\bibitem{martins_evolution_2013}
Martins AB, de~Aguiar MAM, Bar-Yam Y.
\newblock Evolution and stability of ring species.
\newblock Proceedings of the National Academy of Sciences.
  2013;110(13):5080--5084.

\bibitem{Hamilton1963}
Hamilton TH, Rubinoff I.
\newblock {Isolation, Endemism, and Multiplication of Species in the Darwin
  Finches}.
\newblock Evolution. 1963;17(4):388.

\bibitem{Gascuel2016}
Gascuel F, Laroche F, Bonnet-Lebrun AS, Rodrigues ASL.
\newblock {The effects of archipelago spatial structure on island diversity and
  endemism: predictions from a spatially-structured neutral model}.
\newblock Evolution. 2016;70(11):2657--2666.

\bibitem{Barraclough2000}
Barraclough TG, Vogler AP.
\newblock {Detecting the Geographical Pattern of Speciation from Species-Level
  Phylogenies.}
\newblock The American Naturalist. 2000;155(4):419--434.

\bibitem{yamaguchi2021}
Yamaguchi R, Iwasa Y, Tachiki Y.
\newblock Recurrent speciation rates on islands decline with species number.
\newblock Proceedings of the Royal Society B: Biological Sciences.
  2021;288(1949):20210255.

\end{thebibliography}


\begin{thebibliography}{2}%
\makeatletter
\providecommand \@ifxundefined [1]{%
 \@ifx{#1\undefined}
}%
\providecommand \@ifnum [1]{%
 \ifnum #1\expandafter \@firstoftwo
 \else \expandafter \@secondoftwo
 \fi
}%
\providecommand \@ifx [1]{%
 \ifx #1\expandafter \@firstoftwo
 \else \expandafter \@secondoftwo
 \fi
}%
\providecommand \natexlab [1]{#1}%
\providecommand \enquote  [1]{``#1''}%
\providecommand \bibnamefont  [1]{#1}%
\providecommand \bibfnamefont [1]{#1}%
\providecommand \citenamefont [1]{#1}%
\providecommand \href@noop [0]{\@secondoftwo}%
\providecommand \href [0]{\begingroup \@sanitize@url \@href}%
\providecommand \@href[1]{\@@startlink{#1}\@@href}%
\providecommand \@@href[1]{\endgroup#1\@@endlink}%
\providecommand \@sanitize@url [0]{\catcode `\\12\catcode `\$12\catcode
  `\&12\catcode `\#12\catcode `\^12\catcode `\_12\catcode `\%12\relax}%
\providecommand \@@startlink[1]{}%
\providecommand \@@endlink[0]{}%
\providecommand \url  [0]{\begingroup\@sanitize@url \@url }%
\providecommand \@url [1]{\endgroup\@href {#1}{\urlprefix }}%
\providecommand \urlprefix  [0]{URL }%
\providecommand \Eprint [0]{\href }%
\providecommand \doibase [0]{http://dx.doi.org/}%
\providecommand \selectlanguage [0]{\@gobble}%
\providecommand \bibinfo  [0]{\@secondoftwo}%
\providecommand \bibfield  [0]{\@secondoftwo}%
\providecommand \translation [1]{[#1]}%
\providecommand \BibitemOpen [0]{}%
\providecommand \bibitemStop [0]{}%
\providecommand \bibitemNoStop [0]{.\EOS\space}%
\providecommand \EOS [0]{\spacefactor3000\relax}%
\providecommand \BibitemShut  [1]{\csname bibitem#1\endcsname}%
\let\auto@bib@innerbib\@empty
\bibitem [{\citenamefont {Higgs}\ and\ \citenamefont
  {Derrida}(1991)}]{higgs_stochastic_1991}%
  \BibitemOpen
  \bibfield  {author} {\bibinfo {author} {\bibfnamefont {P.~G.}\ \bibnamefont
  {Higgs}}\ and\ \bibinfo {author} {\bibfnamefont {B.}~\bibnamefont
  {Derrida}},\ }\href {\doibase 10.1088/0305-4470/24/17/005} {\bibfield
  {journal} {\bibinfo  {journal} {Journal of Physics A: Mathematical and
  General}\ }\textbf {\bibinfo {volume} {24}},\ \bibinfo {pages} {L985}
  (\bibinfo {year} {1991})}\BibitemShut {NoStop}%
\bibitem [{\citenamefont {Manzo}\ and\ \citenamefont
  {Peliti}(1994)}]{Manzo_Peliti_1994}%
  \BibitemOpen
  \bibfield  {author} {\bibinfo {author} {\bibfnamefont {F.}~\bibnamefont
  {Manzo}}\ and\ \bibinfo {author} {\bibfnamefont {L.}~\bibnamefont {Peliti}},\
  }\href@noop {} {\bibfield  {journal} {\bibinfo  {journal} {Journal of Physics
  A: Mathematical and General}\ }\textbf {\bibinfo {volume} {27}},\ \bibinfo
  {pages} {7079} (\bibinfo {year} {1994})}\BibitemShut {NoStop}%
\end{thebibliography}%

\makeatletter\@input{file1.tex}\makeatother
\end{document}


\title{ Supplementary Material for \\Diversity patterns and speciation processes in \\a two-island system with continuous migration}

 \author{Débora Princepe}
 \author{Simone Czarnobai}
 \author{Thiago M. Pradella}
 \author{Rodrigo A. Caetano}
 \author{Flavia M. D. Marquitti}
 \author{Marcus A.M. de Aguiar}
 \author{Sabrina B. L. Araujo}

 \maketitle

\section{Supplementary Theory}

\subsection{One island dynamics without mating restrictions}
\label{app1}

To understand how the similarity between pairs of individuals changes through generations, we follow \cite{higgs_stochastic_1991} and consider first an asexual population where each individual $\alpha$ has a single parent $P(\alpha)$ in the previous generation. The allele $S_i^\alpha$ will be equal to $S_i^{P(\alpha)}$ with probability $\frac{1}{2}(1+e^{-2\mu})$ and $-S_i^{P(\alpha)}$ with probability $\frac{1}{2}(1-e^{-2\mu})$, then the expected value is $E(S_i^\alpha) = e^{-2\mu} S_i^{P(\alpha)}$. For independent genes, the expected value of the similarity between $\alpha$ and $\beta$ is, therefore,
 \begin{equation}
 E(q^{\alpha \beta}) = e^{-4\mu} q^{P(\alpha) P(\beta)}.
 \label{aveqp}
 \end{equation}
 In sexual populations $\alpha$ and $\beta$ have two parents each, $P_1(\alpha)$, $P_2(\alpha)$ and $P_1(\beta)$, $P_2(\beta)$. Since each one inherits (on average) half the alleles from each parent, we find
\begin{eqnarray}
	E(S_i^\alpha) &=& \frac{e^{-2\mu}}{2}(S_i^{P_1(\alpha)} + S_i^{P_2(\alpha)}). 
	\label{avesexp}
\end{eqnarray}
Using the Eq. 1 of the main text, it follows that, on the average, the similarity between $\alpha$ and $\beta$ is
\begin{equation}
q^{\alpha \beta} = \frac{e^{-4\mu}}{4} \left( q^{P_1(\alpha) P_1(\beta)} +  q^{P_2(\alpha) P_1(\beta)} 
+ q^{P_1(\alpha) P_2(\beta)}  +   q^{P_2(\alpha) P_2(\beta)}  \right)
\label{hdupdatep}
\end{equation}
with $q^{\alpha \alpha} \equiv 1$. In the limit of infinitely many genes, $B \rightarrow \infty$, this expression becomes exact and the entire dynamics can be obtained by simply updating the similarity matrix. 

The dynamics of the average similarity can be derived as follows: the probability that $\alpha$ and $\beta$ have one parent in common is $4/M$ (the four possibilities are: $P_1(\alpha) = P_1(\beta) $, $P_1(\alpha)=  P_2(\beta)$, $P_2(\alpha) = P_1(\beta)$ and $P_2(\alpha)= P_2(\beta)$). In this case, the average similarity between $\alpha$ and $\beta$ would be (see Eq. \ref{hdupdatep})
\begin{equation}
	\langle q^{\alpha \beta} \rangle = \frac{e^{-4\mu}}{4}(1+ 3 q),
\end{equation}
where $q$ is the average similarity in the previous generation. If, on the other hand, $\alpha$ and $\beta$ do not share a parent, which happens with probability $1-4/M$, then
\begin{equation}
	\langle q^{\alpha \beta} \rangle =  e^{-4\mu} q.
\end{equation}
Therefore, at generation $t+1$ we get
\begin{equation}
	q_{t+1} = \frac{4}{M}  \frac{e^{-4\mu}}{4}(1+ 3 q_t) + \left(1-\frac{4}{M}\right)  e^{-4\mu} q_t
\end{equation}
that simplifies to
\begin{equation}
	q_{t+1} = e^{-4\mu} \left[ \left(1-\frac{1}{M}\right)q_t + \frac{1}{M}  \right].
	\label{dynsex}
\end{equation}
Setting $q_{t+1}=q_t$ we find the equilibrium at 
\begin{equation}
	q_0 \approx \frac{1}{1+4\mu M}.
	\label{q0}
\end{equation}
The approximation holds for $\mu$ and $1/M $ much smaller than one, which is always the case for real populations. Subtracting ${q}_t$ from both sides of Eq. \ref{dynsex} and approximating ${q}_{t+1} - {q}_t = \dot{q}$, we get the differential equation
\begin{equation}
	\dot{q} = \frac{1}{M}\left(1-\frac{q}{q_0}\right)
	\label{dyn0}
\end{equation}
whose solution is
\begin{equation}
	q(t) = q_0 + (1-q_0) e^{-t/Mq_0}.
	\label{sol1}
\end{equation}

When imposing the genetic restriction to mating, the time $\tau$ to sympatric speciation can be estimated as the time $q$ takes to reach $q_{min}$. Setting $q(\tau) = q_{min}$ we find
\begin{equation}
	\tau = M q_0 \log \left(\frac{1-q_0}{q_{min}-q_0}\right).
\end{equation}

The population can also be characterized by the distribution of genetic distances $D^{\alpha \beta}$, defined as the number of loci bearing different alleles. It is easy to see that $q^{\alpha \beta} = 1 - 2 D^{\alpha \beta}/B$. In this case mating is allowed only if the genomes differ by at most $G$ loci, or $D^{\alpha \beta}\leq G$. The relation between $G$ and $q_{min}$ is given by 
\begin{equation}
q_{min} = 1-2G/B.   
\end{equation}
In our simulations, we adopted $q_{min} =0.9$, equivalent to $G/B=0.05$.

\subsection{Two islands dynamics}

From the Eq. 3 of the main text, derived in \cite{Manzo_Peliti_1994}, we obtain the dynamics of the similarities $q$ and $p$. For $\epsilon$,  $\mu$ and $1/M$ all much smaller than 1, the equations can be approximated by
\begin{eqnarray}
	q_{t+1} &=& (1-2\epsilon - 4 \mu - M^{-1})q_t + M^{-1} + 2 \epsilon p_t \nonumber \\
	p_{t+1} &=& 2 \epsilon q_t + (1- 2\epsilon - 4\mu)  p_t. 
	\label{qp_sys}
\end{eqnarray}
Using the approximations $q_{t+1}-q_{t}=\dot{q}$ and $p_{t+1}-p_{t}=\dot{p}$, we obtain the differential equations
\begin{eqnarray}
	\dot{q} &=& -(2\epsilon + 4 \mu + M^{-1})q + M^{-1} + 2 \epsilon p \nonumber \\
	\dot{p} &=& 2 \epsilon q - (2\epsilon + 4\mu)  p.
\label{dyn1}	
\end{eqnarray}
We obtain the solutions
\begin{eqnarray}
    q(t) &=& \left(a_1 -\frac{a_1}\lambda_{+}-\frac{2 \sigma}{\sqrt{\Delta}}\right)e^{-\lambda_+ t/M}+2\sigma\left( -a_2 +\frac{a_2}{\lambda_-} +\frac{1}{\sqrt{\Delta}}\right)e^{-\lambda_- t/M} +q(0)\nonumber \\ \\
    p(t) &=&\left( \frac{1-\sqrt{\Delta}}{4\sigma}\right) \left(a_1 -\frac{a_1}\lambda_{+}-\frac{2 \sigma}{\sqrt{\Delta}}\right)e^{-\lambda_+ t/M}+ \nonumber \\
    & &\left( \frac{1+\sqrt{\Delta}}{2}\right)\left( -a_2 +\frac{a_2}{\lambda_-} +\frac{1}{\sqrt{\Delta}}\right)e^{-\lambda_- t/M}+p(0)
\end{eqnarray}
where $q(0)=p(0)=1$ and
\begin{eqnarray}
    & a_1=\frac{1+\sqrt{\Delta}}{2\sqrt{\Delta}}, \quad 
    a_2=\frac{1-\sqrt{\Delta}}{4\sigma\sqrt{\Delta}}, \quad 
    \lambda_\pm=2\sigma + \nu + \frac{1}{2} \pm \frac{1}{2}\sqrt{\Delta}, \quad& \Delta=16 \sigma^2 +1.
\end{eqnarray}

In the limit $\sigma \rightarrow 0$, the Eq. \ref{sol1} for $q(t)$ is recovered, and $p(t)$ is simply given by $p(t)=e^{- \nu t/M}$. Considering two isolated islands without migration, the time for the speciation due to geographical isolation ($\tau_a$) is approximated by $p(\tau_a)=e^{- \nu \tau_a/M} = q_{min}$. The time for allopatry can now be derived:
\begin{eqnarray}
    \tau_a=\frac{1}{4\mu} \log {\frac{1}{q_{min}}}.
    \label{eq:tallop}
\end{eqnarray}

%
%
%

\subsection{Effective number of migrants}

Here we calculate the number of migrants that effectively establish a sub-population in the arrival island, $\sigma_{eff}$. The probability that $\sigma$, of a total of $M$ individuals, migrate to the other island is 
\begin{equation}
    P_M(\sigma) = \binom{M}{\sigma} \epsilon^\sigma (1-\epsilon)^{M-\sigma}
\end{equation}
where $\epsilon$ is migration probability per individual. These individuals will only give rise to a community if they reproduce and leave offspring to the next generation. For that, they have to be selected for reproduction (from the whole population $M$ in $M$ trials) and then select a mating partner among the migrants.  

The probability of {\it not} selecting one among the $\sigma$ migrants in one attempt is $1 - \sigma/M$. The probability of not selecting one among the $\sigma$ migrants in $M$ attempts is $(1 - \sigma/M)^M \approx e^{-\sigma}$. Therefore the probability of selecting one migrant in $M$ attempts is $(1-e^{-\sigma})$. Similarly, the probability that the reproducing migrant selects another migrant as mating partner is $(1-e^{-(\sigma-1)})$.
Therefore, we can define an effective number of migrants (corresponding to successful migrants) as
\begin{eqnarray}
    \sigma_{eff} = \frac{1}{N}\sum_{k=1}^{N} \sum_{\sigma=2}^{m_k} \sigma P_{m_k}(\sigma) (1-e^{-\sigma}) (1-e^{-(\sigma-1)}) 
\end{eqnarray}
where $N$ is the number of species in the island where the migrants were born and $m_k$ is the abundance of the k-th species. Because the migrants can belong to different species, we have constrained reproduction among individuals of the same species of migrants and averaged over all species. We can simplify this expression using the approximation that all species have the same number of individuals: $m=(q_{min}^{-1} - 1)/4\mu$ and $N=M/m$. We obtain
\begin{eqnarray}
    \sigma_{eff} = \sum_{\sigma=2}^m \sigma \binom{m}{\sigma} \epsilon^\sigma (1-\epsilon)^{m-\sigma} (1-e^{-\sigma}) (1-e^{-(\sigma-1)}).
\end{eqnarray}
If the probability $P_m(\sigma)$ is peaked on the average value $\bar\sigma = M\epsilon$, we can further approximate this expression by
\begin{eqnarray}
    \sigma_{eff} = \bar\sigma (1-e^{-\bar\sigma}) (1-e^{-(\bar\sigma-1)})
\end{eqnarray}
which recovers $\sigma_{eff} = \bar\sigma$ for $\bar\sigma \gg 1$. We considered the effective number of migrants for the calculations of the expected number of species $N$ and $N_T$ (Fig. 3 a and b in the main text).
\newpage
\section{Supplementary Results}

\subsection{Number of simulations}

\begin{figure}[!h]
    \includegraphics[width=0.7\linewidth]{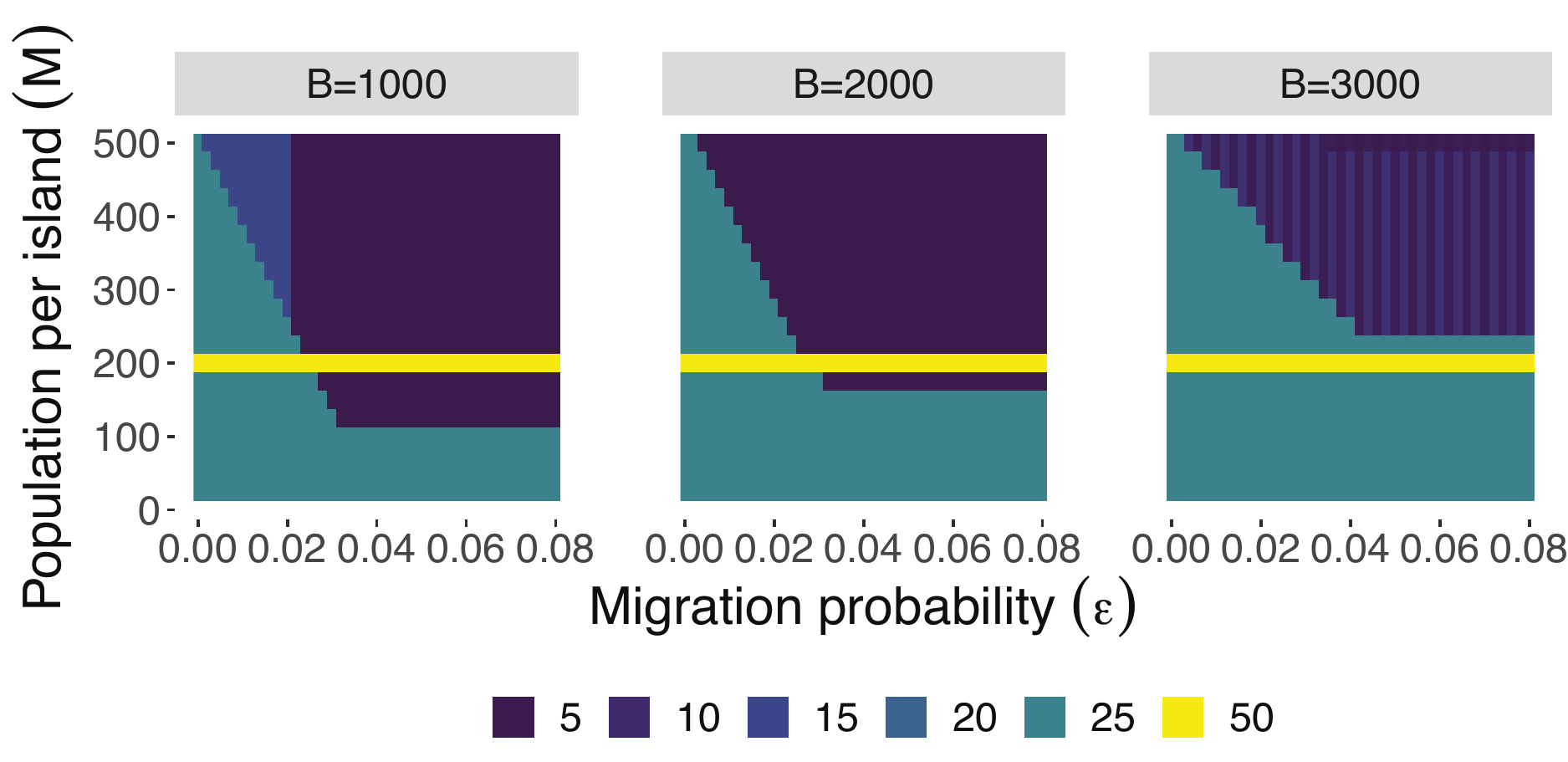}
  \caption{ Number of simulations (color scale) performed to compose Figure 2 of the main text.}
  \label{fig:Samples}
\end{figure}

Figure \ref{fig:Samples} shows the number of simulations performed for each genome size, population size, and migration probability to compose Figure 2 of the main text. For parameter values resulting in a small number of species ($\approx1$), we could make fewer simulations to calculate the average value since fluctuations were small. This was convenient for simulations of finite genomes that demand high computational time.
Notice that few realizations are sufficient above the migration intensity for which the number of species collapses to a single one shared. For the analysis with fixed population size $M=200$ (Fig. 3 and 4 of the main text), we ran 50 simulations in all cases.

\subsection{Evolution of diversity for $B=1000$, $2000$, and $3000$}

\begin{figure}[h]
    \includegraphics[width=0.9\linewidth]{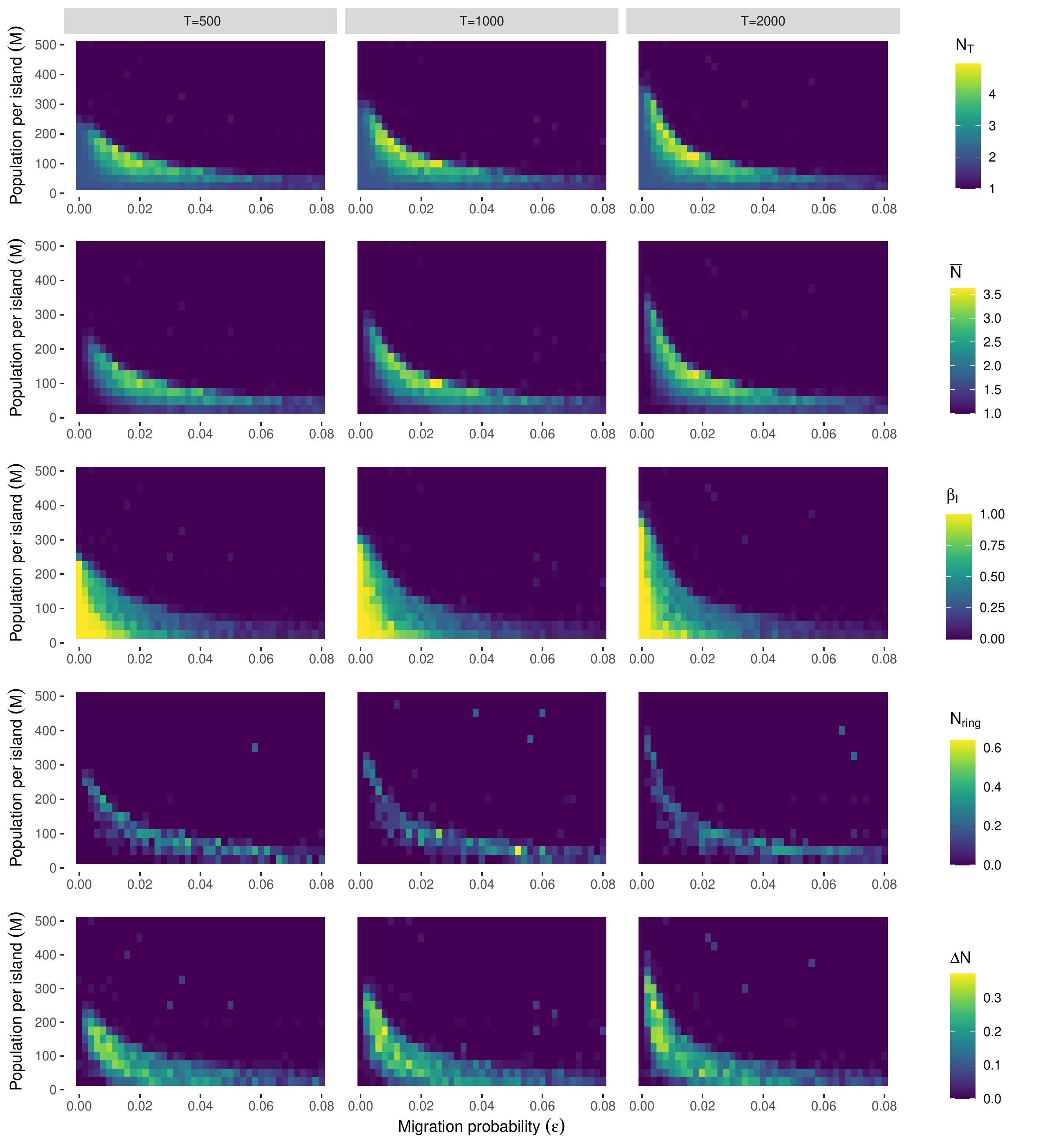}
  \caption{Diversity patterns of the two-island system as a function of population size ($M$) and migration probability ($\epsilon$) for $B=1000$ at $T=500$ (left column), $1000$ (middle) and $2000$ (right column) generations. }
  \label{fig:B1000}
\end{figure}

\begin{figure}[!h]
    \includegraphics[width=0.9\linewidth]{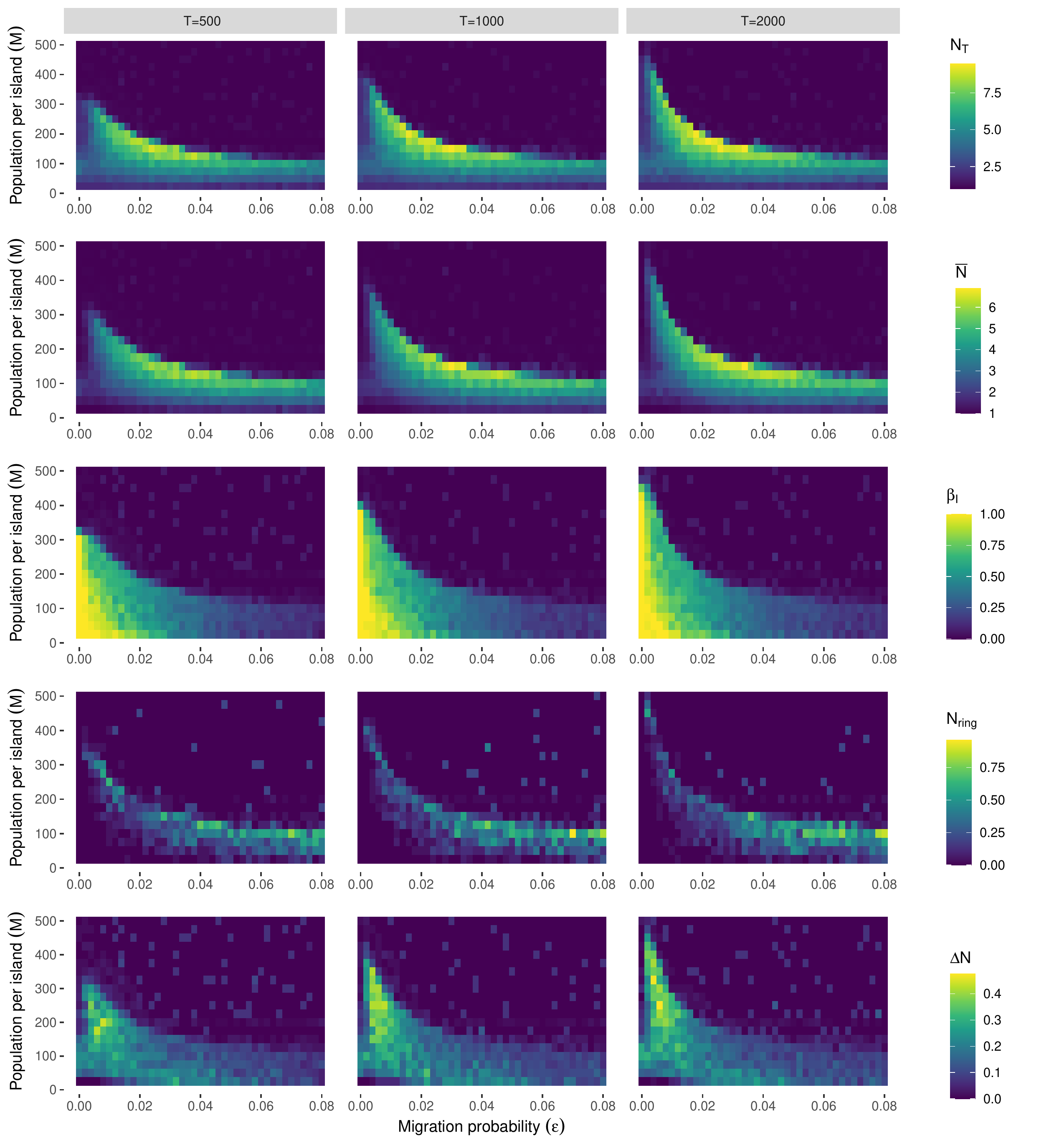}
  \caption{Diversity patterns of the two-island system as a function of population size ($M$) and migration probability ($\epsilon$) for $B=2000$ at $T=500$ (left column), $1000$ (middle) and $2000$ (right column) generations.}
  \label{fig:B2000}
\end{figure}

\begin{figure}[!h]
    \includegraphics[width=\linewidth]{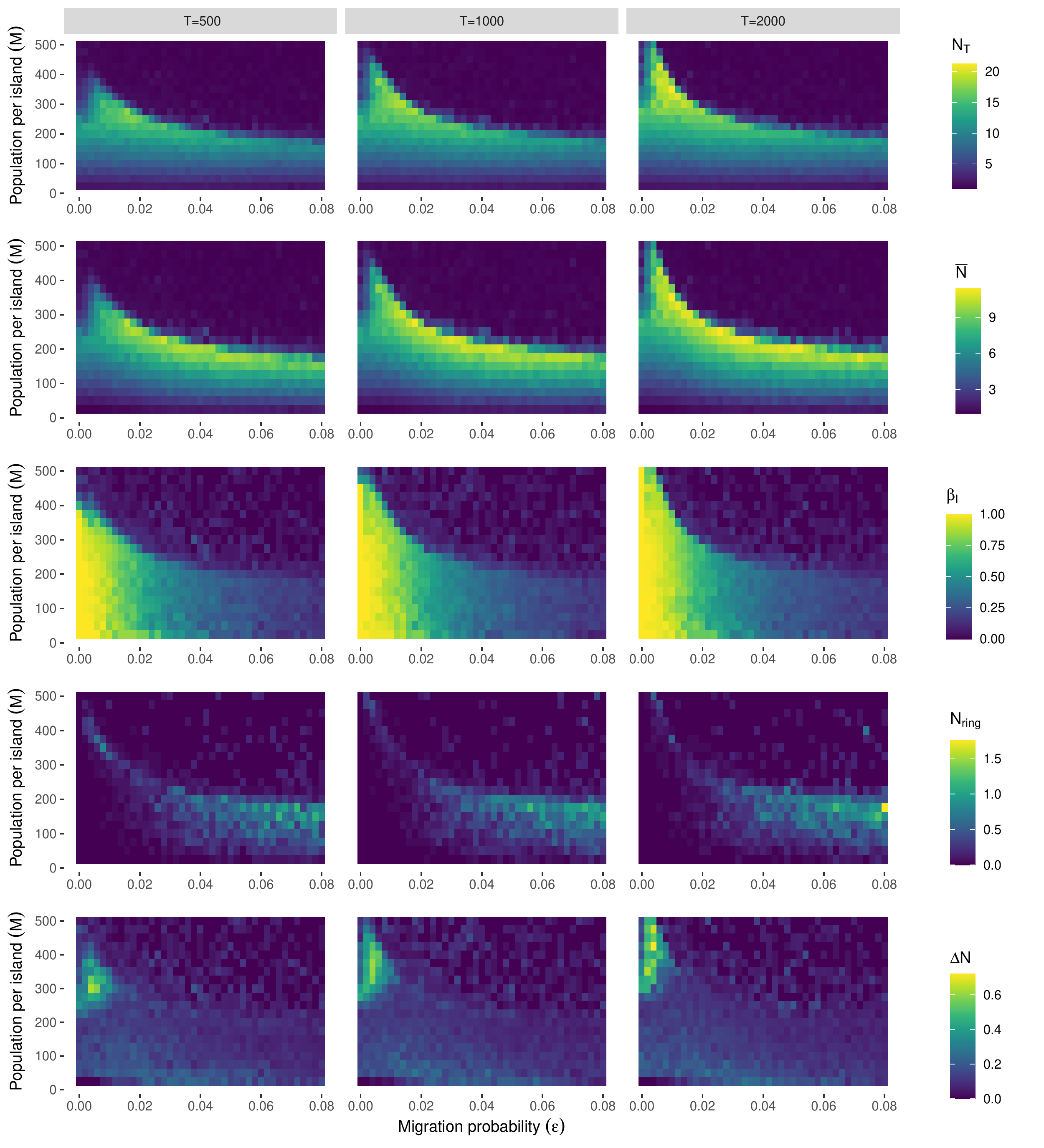}
  \caption{Diversity patterns of the two-island system as a function of population size ($M$) and migration probability ($\epsilon$) for $B=3000$ at $T=500$ (left column), $1000$ (middle) and $2000$ (right column) generations.}
  \label{fig:B3000}
\end{figure}

Here we present additional results to support our analysis in the manuscript. Figures \ref{fig:B1000}, \ref{fig:B2000} and \ref{fig:B3000} show the patterns of diversity for the genome sizes $B=1000$, $2000$, and $3000$ respectively, similarly to Figure 2 in the main text, in different time steps: $T=500$, $1500$ and $2000$ generations. In each figure, the columns represent the time of the observation. From top to bottom, the plots represent the average values of the total number of species in the insular system ($N_T$), the number of species per island ($\bar{N}=(N_{1}+N_{2})/2$), the ratio of the total number of species that are exclusive to an island ($\beta_I$), the number of ring-like species ($N_{ring}$), and the asymmetry in species richness ($\Delta N= |N_{1}-N_{2}|/\bar{N}$). The numbers of simulations utilized follow the indicated in Figure \ref{fig:Samples}. The diversity patterns were consistent over time, but larger populations took longer to speciate in allopatry or with migration, i.e., the time to speciation depends on $M$ for finite genomes. 

\begin{figure}[h]
    \centering
    \begin{minipage}{0.32\textwidth}
    \centering
    { $B=1000$}
    \includegraphics[width=\linewidth]{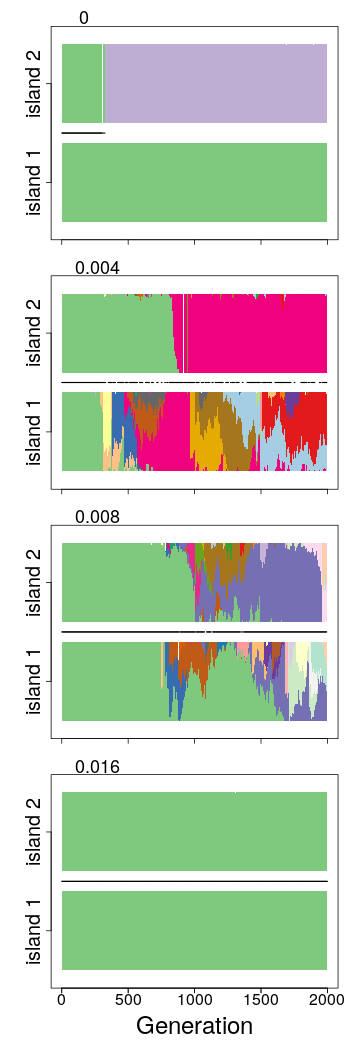}
    \end{minipage}
  \hspace*{\fill}
      \begin{minipage}{0.32\textwidth}
    \centering
    {$B=2000$}
    \includegraphics[width=\linewidth]{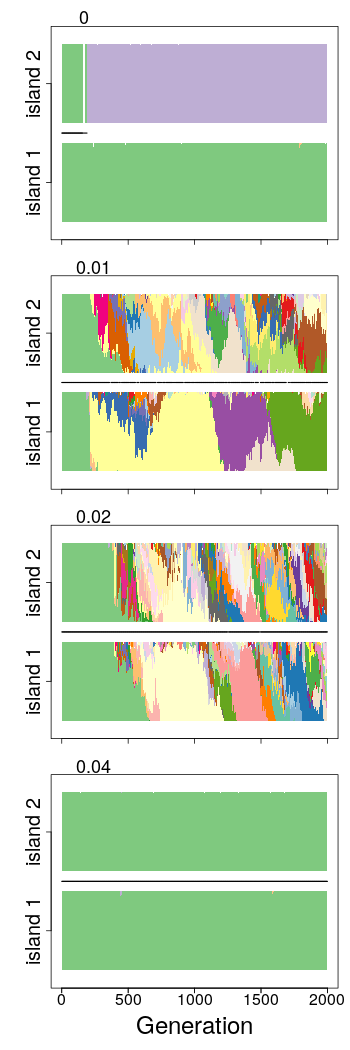}
    \end{minipage}
  \hspace*{\fill} 
    \begin{minipage}{0.32\textwidth}
    \centering
    {$B=3000$}
    \includegraphics[width=\linewidth]{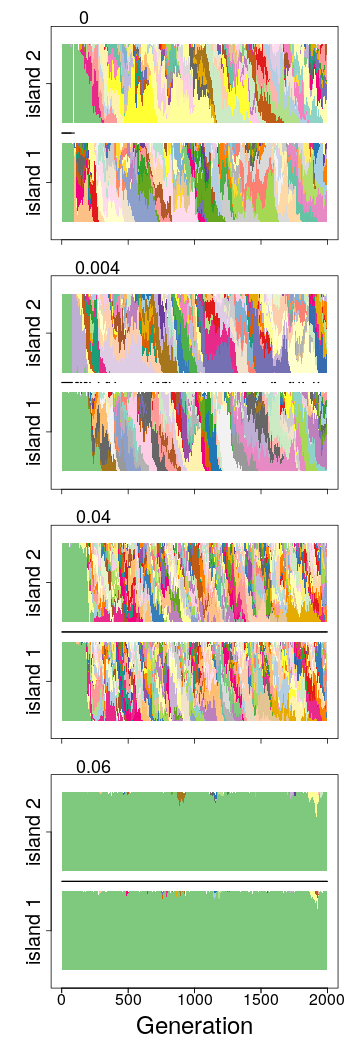}
    \end{minipage}
  \caption{Distribution of species per island over time for different migration probabilities (indicated in each panel) and fixed population size $M=200$, for $B=1000$ (left column), $2000$ (middle) and $3000$ (left column). Each panel has two horizontal bars corresponding to each island. The colors identify species, and their vertical amplitude represents the species abundance. The horizontal black line between the bars indicates the intervals when the islands shared at least one common species.} 
  \label{fig:Temporal}
\end{figure}

Figure \ref{fig:Temporal} depicts the dynamics in the two islands for some values of migration probability and fixed population size $M=200$. The colors identify species, and the height is proportional to the species abundance. In allopatry ($\epsilon=0$), we observe that sympatric speciation occurred only for $B=3000$. Also, differentiation between islands took longer the shorter the genome size. Migration induced speciation up to a critical value of $\epsilon$ that collapsed the populations to a single shared species. Longer genomes supported higher migration flux before collapsing. 

\clearpage

\subsection{Number of exclusive species}
\label{SM-exc}

\begin{figure}[ht]
    \includegraphics[width=0.45 \linewidth]{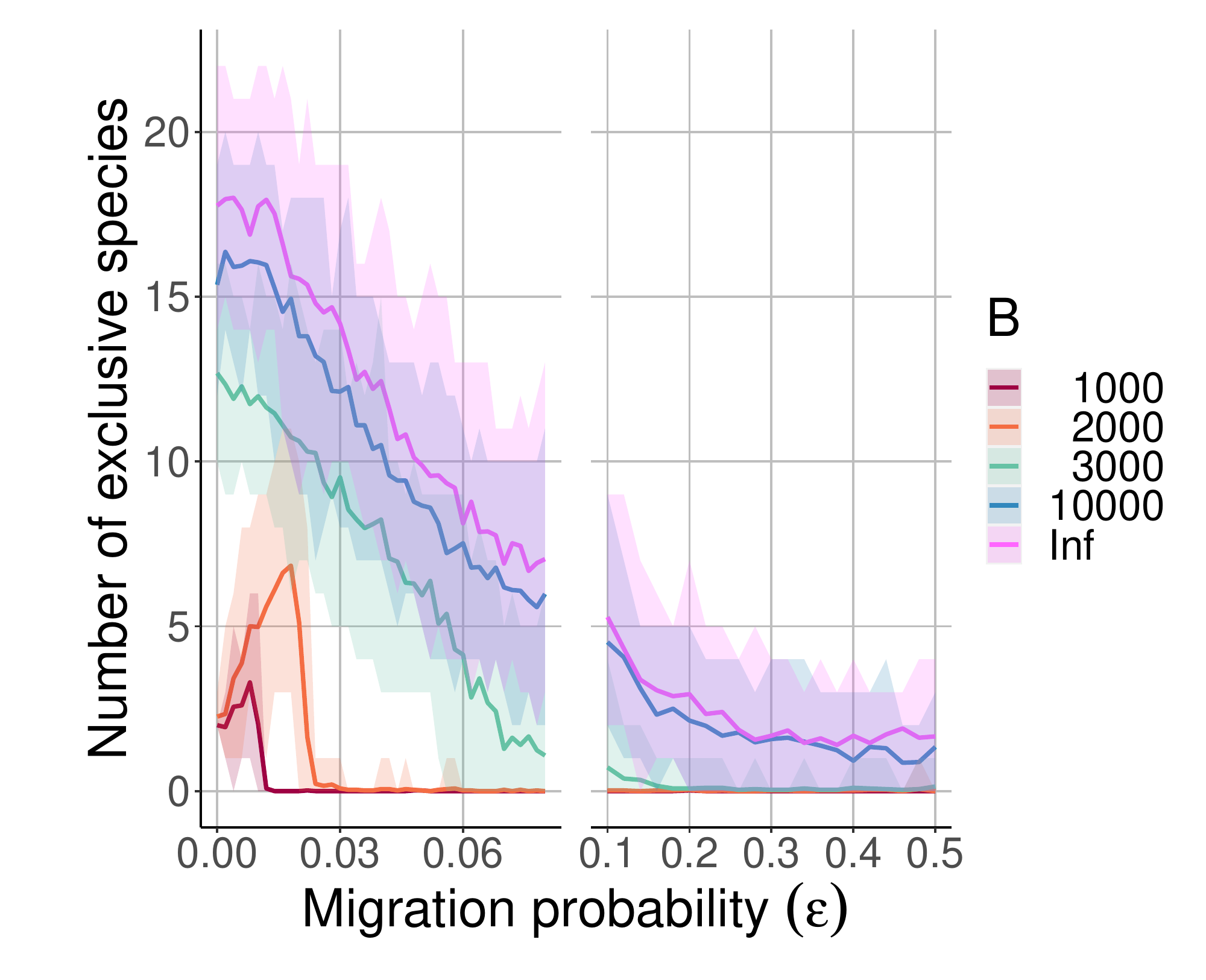} 
  \caption{Number of exclusive species in the system, $K_1+K_2$.     }
  \label{fig:exclusive}
\end{figure}

Figure \ref{fig:exclusive} depicts the number of exclusive (endemic) species in the insular system for varied genome sizes, complimentary to Figures 3 and 4 of the main text. The number of exclusive species in each island can be calculated as follows: calling $K_i$ the number of endemic species to the island $i$ and $c$ the number of common species, then $N_{1}=K_1+c$, $N_{2}=K_2+c$, and $N_T=K_1+K_2+c$. Then $c=N_{1}+N_{2}-N_T$, and $K_1$ and $K_2$ are trivially calculated. The beta diversity index is given by the ratio $(K_1+K_2)/N_T$.

\subsection{Trade-off between genome length ($B$), minimal similarity for reproduction ($q_{min}$), mutation rate ($\mu$) and population size ($M$) }
\label{SM-tradeoff}

\begin{figure}[ht]
    \includegraphics[width=0.45 \linewidth]{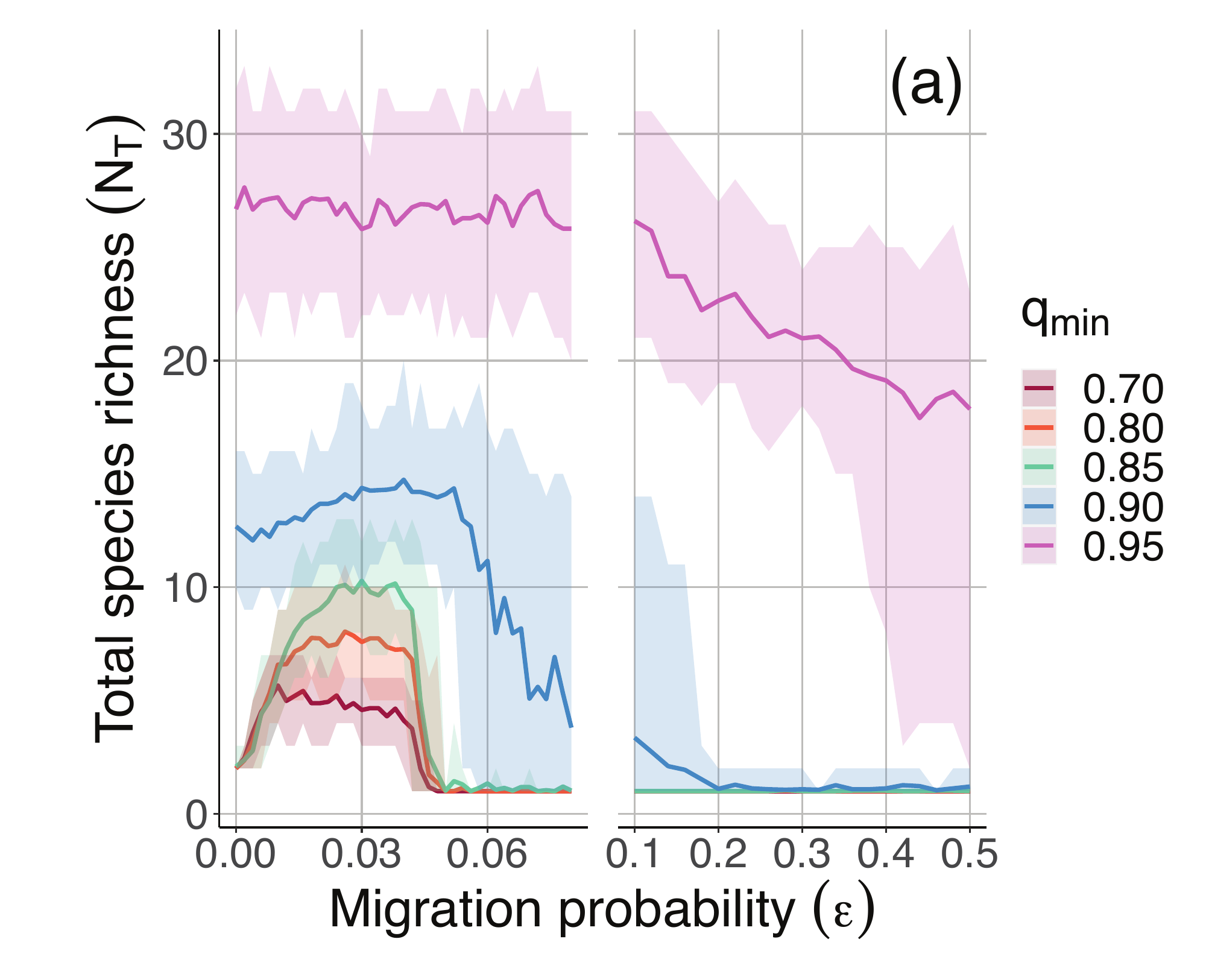} 
    \includegraphics[width=0.45\linewidth]{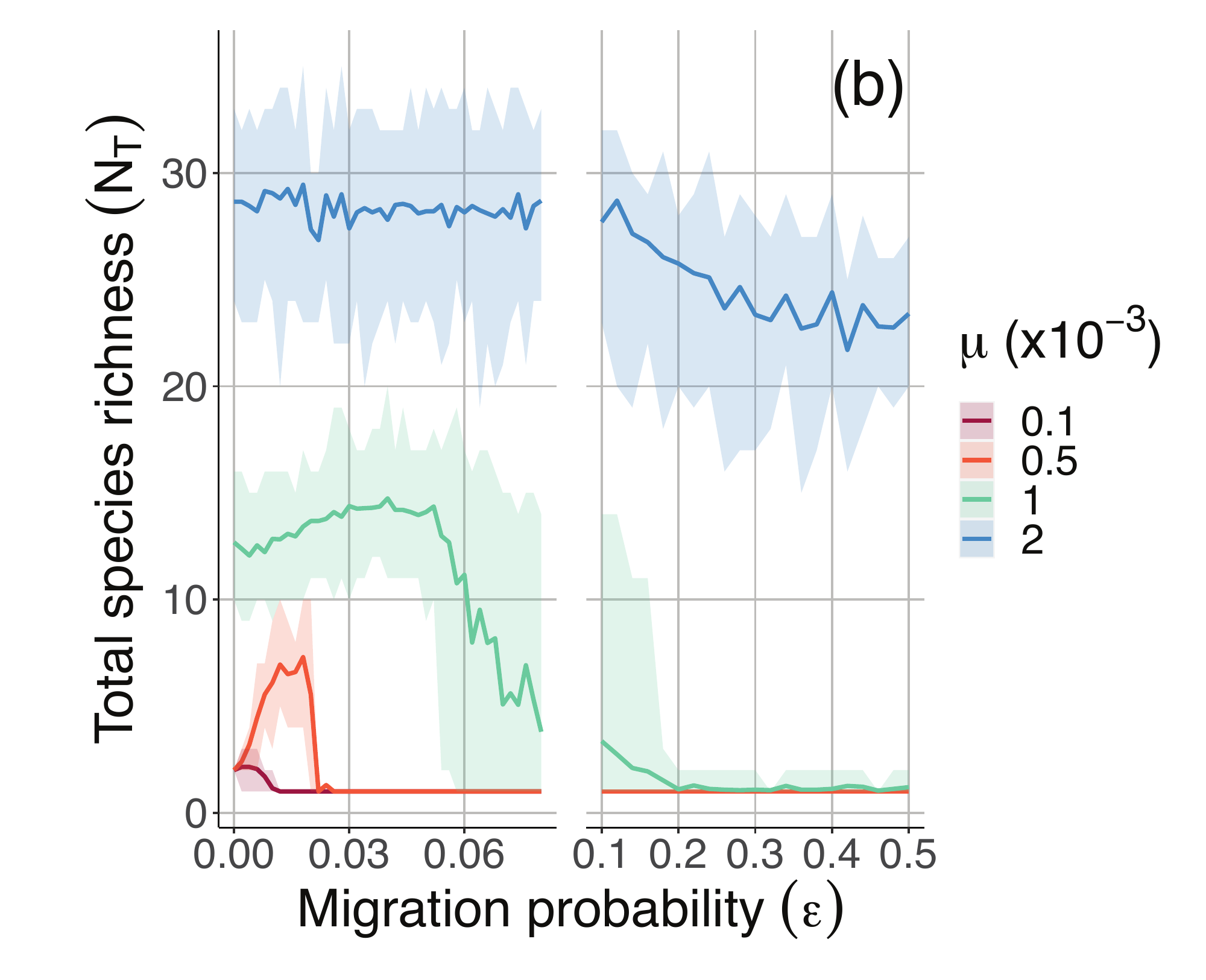}
  \caption{ Effect of the minimal genetic similarity $q_{min}$ (a) and mutation rate $\mu$ (b) on the total species richness. The other simulation parameters, unless varied in the graph, were:  $B=3000$, $M=200$, $\mu=0.001$, and $q_{min} =0.9$.}
  \label{fig:qmin}
\end{figure}

In general, by decreasing the minimal genetic similarity for reproduction $q_{min}$, the total number of species decreases, following the trend of the number of species per island, $N$ (Eq. 5 of the main text). Figure \ref{fig:qmin}a shows that the qualitative patterns obtained when $B$ increases while keeping the other parameters fixed (Fig. 3b) can be recovered by decreasing $q_{min}$. 
A similar scaling of effects occurs for the mutation rate: the patterns obtained by decreasing $B$ (Fig. 3b) are recovered by decreasing the mutation rate (Fig. \ref{fig:qmin}b).



\begin{figure}[ht]
    \includegraphics[width=0.45\linewidth]{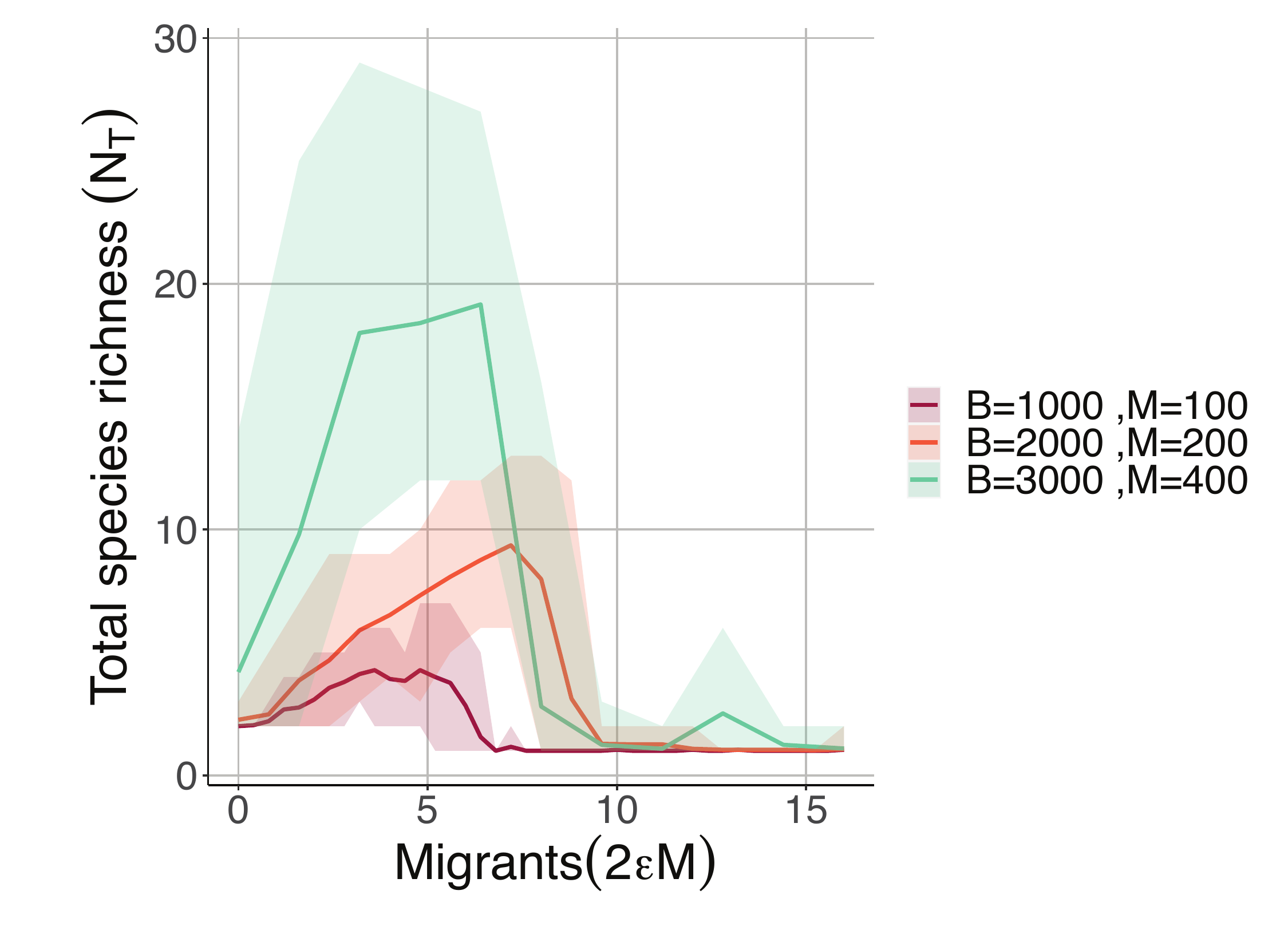}
  \caption{Selected sets of $B$ and $M$ from Figure 2 of the main text showing the behavior of $N_T$ with the number of migrants.}
  \label{fig:fig2rev}
\end{figure}

Furthermore, we expect the observed effects to scale with genome length and population size, as shown in Figure 2. In Figure \ref{fig:fig2rev}, we choose sets of $B$ and $M$ from the map in Figure 2 that demonstrates the similar behavior of the total number of species with the migration rate (here plotted as a function of the number of migrants for convenience), i.e., the increase in richness when there are few migrants, followed by the collapse to a single shared species for a high migration intensity. Therefore, although we were limited to simulating small population sizes and genome lengths due to the computational time, we expect similar effects for larger populations with larger values $B$.

\subsection{Null test of asymmetry}
\label{SM-asy}
\setlength\parskip{0em}

The asymmetry in species richness, calculated by the absolute value of the difference between the number of species in each island, $|N_1-N_2|$, is equivalent to the modulus of the difference between the number of exclusive (endemic) species, $|K_1-K_2|$. To verify if the asymmetry observed in our simulations was significant, we compared it to a null model where the exclusive species were randomly distributed between the islands with equal probability. We run the null model for each simulation output using the number of exclusive species from the simulation. In each case, we ensured that at least one species inhabited each island: if the islands shared at least one species, $K_1+K_2$ species were randomly distributed between the islands, otherwise only $K_1+K_2-2$ species were distributed, since one species is placed at each island at the start. To compare the different genome sizes, we adopted the asymmetry normalized by the total number of species, $\Delta N$ (Eq. 7 in the main text).

\begin{figure}[ht]
    \includegraphics[width=\linewidth]{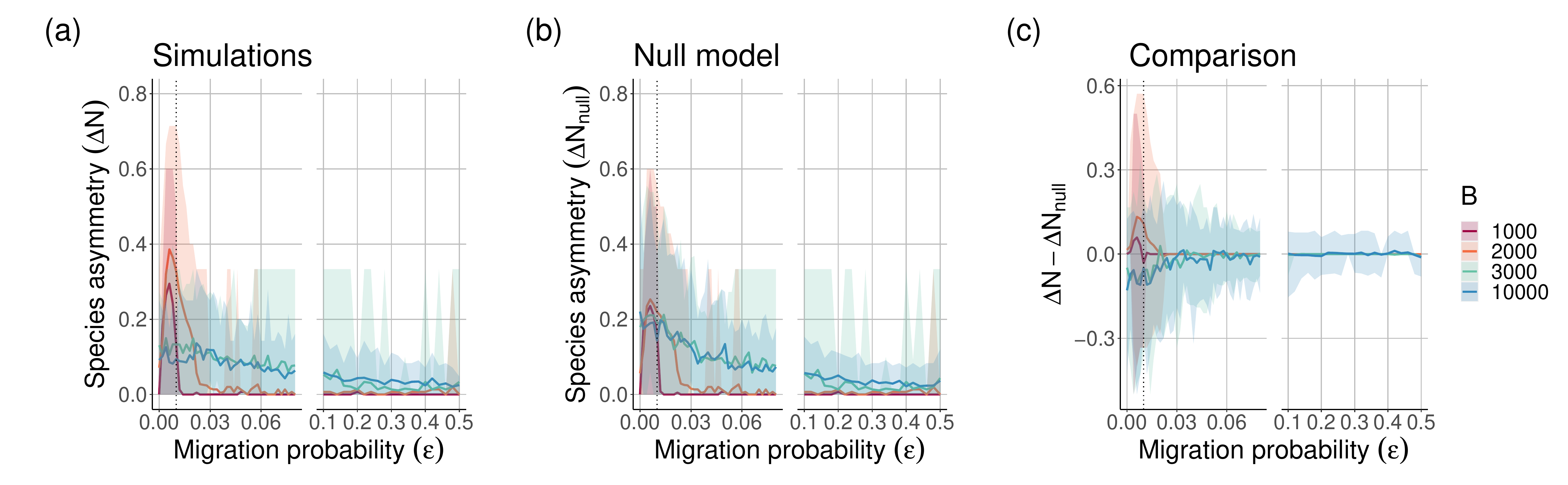}
  \caption{Asymmetry in species richness in the insular system ($\Delta N$) for finite genomes. The solid lines represent the average value of all simulations, and the shadowed areas show a confidence interval of  90\%. (a) Results from our simulations (similar to the plot in the manuscript (Fig. 3c) considering only finite genomes and more data (some averages were calculated over 1000 repetitions). (b) Species asymmetry expected by the null model, where species are randomly distributed over the islands. (c) Comparison between the obtained from simulation and null model ($\Delta N-\Delta N_{null}$) reveals that asymmetry is more expected than by chance (positive values) for low migration probability and small genomes, $B=1000$ and $2000$. The dashed vertical line highlight $\epsilon=0.01$ (same as Fig. \ref{fig:Histograms}).}
  \label{fig:sup_asy}
\end{figure}

Figure \ref{fig:sup_asy} shows that the simulations differed from the random distribution when the migration probability was low ($\epsilon<0.03$). For small genomes ($B= 1000$ and $2000$), the asymmetry was higher than expected by chance, while for large genomes ($B=3000$ and $10000$), it was lower than the expectation at random. The higher asymmetry for small genomes suggests that speciation by founding populations was probably the main speciation mechanism associated: random imbalances in $N_1$ and $N_2$ were enhanced because migrants leaving the island with lower richness were more likely to be genetically compatible and to found a new species in the arrival island. On the other hand, the number of species was higher for larger genomes, decreasing the fluctuations in species richness and the likelihood of founding populations. 

\begin{figure}[!
ht]
    \includegraphics[width=\linewidth]{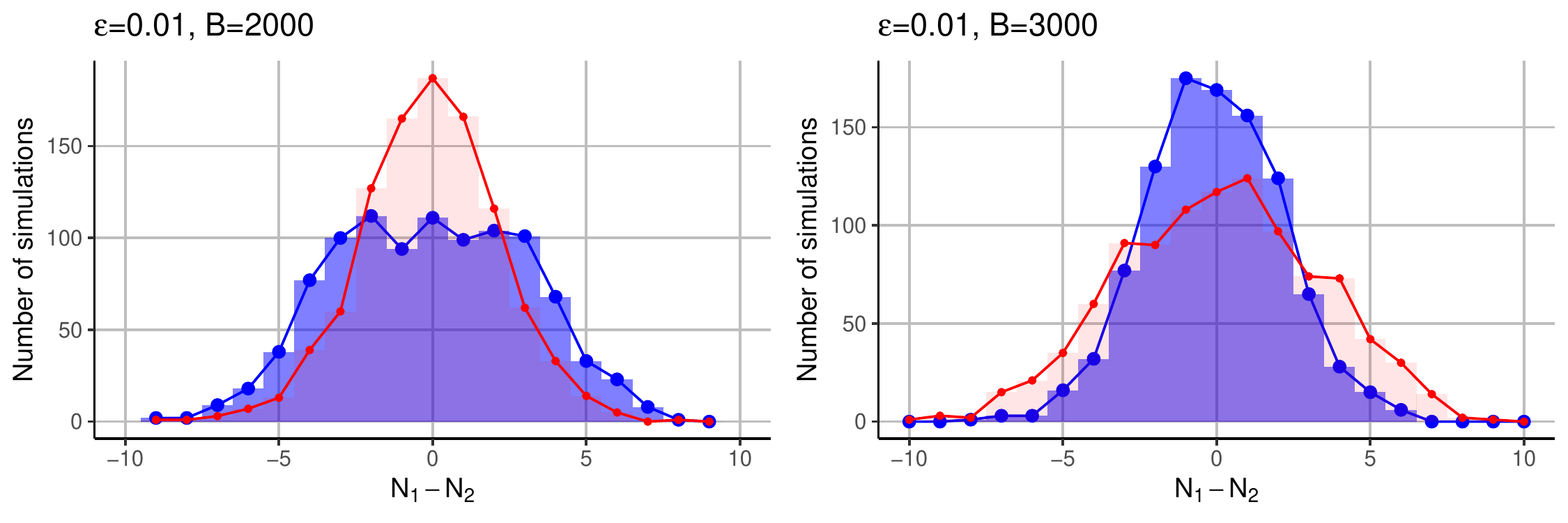}
  \caption{Histograms of the difference between the islands species richness, $N_1 - N_2$, obtained from the simulations (blue) and the null model (red) for 1000 replications when the migration probability is $\epsilon=0.01$. For $B=2000$ (left), the asymmetry is higher than expected by chance, while for $B=3000$ (right) the asymmetry is lower than expected by chance. }
  \label{fig:Histograms}
\end{figure}

The divergence between the simulations and the null model is better observed in Figure \ref{fig:Histograms}, which compares the difference $N_1-N_2$ (here is not the absolute value) for $1000$ realizations with $\epsilon=0.01$ and two values of genome size, $B= 2000$ (left) and $3000$ (right). The distribution of $N_1-N_2$ when $B=2000$ has a plateau for $|N_1-N_2|\leq3$ (blue bars), revealing that perfect symmetry ($N_1-N_2=0$) is not more likely within this range, in opposite to what is expected by chance (red bars). On the other hand, the distribution of $N_1-N_2$ for $B=3000$ (right, blue bars) is narrower than expected by chance (red bars), then perfect symmetry is more likely to occur.

\bibliography{referencias}

\makeatletter\@input{file.tex}\makeatother